\documentclass[12pt]{article}
\usepackage{jheppub}
\usepackage{cancel}
\usepackage{subcaption}
\usepackage{xcolor}
\usepackage{amsmath}
\usepackage{amssymb}
\usepackage{amscd}
\usepackage{dsfont}
\usepackage{enumerate}
\usepackage{enumitem}
\usepackage{amsfonts}
\usepackage{epsfig}
\usepackage{mathtools}
\usepackage{yfonts}
\usepackage{bbold}
\usepackage{breqn}
\usepackage[utf8]{inputenc}
\usepackage[english]{babel}
\usepackage{import}
\usepackage{caption}
\usepackage{multirow}

\def\eqa{\begin{eqnarray}}
\def\eqae{\end{eqnarray}}
\def\eq{\begin{equation}}
\def\eqe{\end{equation}}
\def\be{\begin{equation}}
\def\ee{\end{equation}}
\def\bea{\begin{eqnarray}}
\def\eea{\end{eqnarray}}
\def\ba{\begin{array}}
\def\ea{\end{array}}

\usepackage{braket} 
\newcommand*{\dd}{\mathop{}\!\mathrm{d}} 

\title{Non-local modular flows across deformed null-cuts}
\author{Guan-Cheng Lu, Huajia Wang}
\affiliation{Kavli Institute for Theoretical Sciences, University of Chinese Academy of Sciences, \\ Beijing 100190, China}
\emailAdd{luguancheng22@mails.ucas.ac.cn}
\emailAdd{wanghuajia@ucas.ac.cn} 

\abstract
{Modular flows probe important aspects of the entanglement structures, especially those of QFTs, in a dynamical framework. Despite the expected non-local nature in the general cases, the majority of explicitly understood examples feature local space-time trajectories under modular flows. In this work, we study a particular class of non-local modular flows. They are associated with the relativistic vacuum state and sub-regions whose boundaries lie on a planar null-surface. They satisfy a remarkable algebraic property known as the half-sided modular inclusion, and as a result the modular Hamiltonians are exactly known in terms of the stress tensor operators. To be explicit, we focus on the simplest QFT of a massive or massless free scalar in $1+2$ dimensions. We obtain explicit expressions for the generators. They can be separated into a sum of local and non-local terms showing certain universal pattern. The preservation of von-Neumann algebra under modular flow works in a subtle way for the non-local terms. We derive a differential-integral equation for the finite modular flow, which can be analyzed in perturbation theory of small distance deviating from the entanglement boundary, and re-summation can be performed in appropriate limits. Comparison with the general expectation of modular flows in such limits are discussed.}

\begin{document}
\maketitle
\section{Introduction}\label{sec:intro}
The structure of quantum entanglement has played a key role in revealing deep aspects of quantum field theories (QFTs) and AdS/CFT duality \cite{Maldacena:1997re,Maldacena1999,Witten1998AntideSS} not accessible using conventional tools and techniques. A non-exhaustive list of examples include the order parameter for topological phase \cite{Levin_2006,Kitaev_2006}, the famous Ryu-Takayanagi formula \cite{Ryu2006} and its subsequent generalizations \cite{Hubeny2007,Faulkner_2013,Lewkowycz_2013,Engelhardt_2015} encoding bulk space-time geometry in holography, the characterization of universal monotonous properties of renormalization group flow \cite{Myers2011Jan,Casini:2015woa,Casini:2017vbe,Casini_2023}, and the relation between information theory and energy conditions \cite{PhysRevD.93.024017,HWang2016Sep,HWang2019Sep}, etc.

In spatially extended many-body systems such as QFTs, the notion of quantum entanglement is defined w.r.t a global state $|\psi\rangle$ and a subregion $A$. The key object is the reduced density matrix defined by: 
\be
\rho^\psi_A = \text{Tr}_{\bar{A}} |\psi\rangle \langle \psi |.
\ee
This definition makes good sense for discrete systems or discrete approximation of continuous systems such as QFTs. In the latter cases, additional subtleties arise in the continuous limit concerning the lack of tensor factorized structure for the underlying Hilbert space, as well as the nature of the trace operators. As a result one has to rely on more sophisticated tools from the algebraic QFTs to make sense of the entanglement structures \cite{borchers2000revolutionizing,Sorce_2023,Witten_2018,Witten:2021jzq,Leutheusser:2021qhd,Leutheusser:2021frk,leutheusser2024subregionsubalgebradualityemergencespace}, which has also shed lights on the algebraic aspects of perturbative quantum gravity \cite{Witten_2022,Chandrasekaran_20231,Chandrasekaran_20232,Kudler-Flam:2023qfl,Jensen:2023yxy}.  

Useful quantitative measures can be defined regarding bi-partite entanglement such as the von Neumann entropy, R\'enyi entropy, mutual information, or entanglement negativity \cite{PhysRevA.65.032314}; it can also be extended to quantify multi-partite entanglements \cite{PhysRevD.35.3066}. For bi-partite entanglements, more refined data concerning the entanglement structure are encoded in the so-called modular Hamiltonian, defined from the reduced density matrix via the operator relation: 
\be
\rho^\psi_A = \exp{\left(-H^\psi_A\right)}.
\ee
Being an operator, it naturally encodes much richer information than a single quantitative measure. For example, the commutators between vacuum modular Hamiltonians for intersecting regions can be used to characterizing the topological phases in 2+1 dimensional gapped systems \cite{PhysRevLett.128.176402, PhysRevLett.131.186301}. In the context of AdS/CFT, the modular Hamiltonian has also been instrumental in elucidating subtle aspects related to the bulk/boundary correspondence. In particular, it is believed to be crucial in realizing the entanglement wedge reconstruction, thus manifesting the notion of sub-region duality \cite{Dong_2016,Faulkner:2017vdd,leutheusser2024subregionsubalgebradualityemergencespace}.   

From an operator perspective, the modular Hamiltonian can be most straightforwardly analyzed by dissecting it through the spectral decomposition. In discrete systems such as lattice or tensor models, this may be performed explicitly and with reasonable effectiveness.  In the continuous limit of QFTs, it is more convenient to take an indirect approach and study the dynamics generated by the modular Hamiltonian, which in principle contains the same information as the spectral decomposition. In the Schrodinger picture, this dynamics is characterized by the fact that the defining state $\rho^\psi_A$ is the equilibrium state: 
\be
\rho^\psi_A(s) = e^{iH^\psi_A s} \left(\rho^\psi_A\right) e^{-iH^\psi_A s}=\rho^\psi_A .
\ee
The ``near-equilibrium" dynamics can be described by considering the operator flow in the Heisenberg picture: 
\be
\mathcal{O}^s = e^{iH^\psi_A s} \mathcal{O} e^{-iH^\psi_A s}.
\ee
This defines the ``modular flow" of operators, and $s$ is called the modular parameter.  

 A special class of modular flows are those that generate trajectories localized in space-time, i.e. local operators in space-time remain local after modular flow \footnote{We clarify that the notion of operators localized in space and time should not be confused with that of local operators defined w.r.t. the Hilbert space on a fixed time-slice. The former may spread and become non-local w.r.t. the latter via the Heisenberg equation of motion. }:
\be
\mathcal{O}^s(t,x)=e^{iH^\psi_A s}\mathcal{O}(t,x) e^{-iH^\psi_A s} \propto \mathcal{O}\left(t(s),x(s)\right).
\ee
Modular flows of this nature are called local or geometric modular flows. This marks unusual properties of the modular Hamiltonian that requires some fine-tuning, e.g. in the form of (conformal) isometries of the defining states \cite{Sorce_2024,Chen_2023}. The few explicit examples include the half-plane (spherical) region in the vacuum of QFTs (CFTs). As a result, local modular flows can lead to remarkable implications. In the context of von Neumann algebra, local modular flows play important roles in constructing the type II algebra underlying perturbative quantum gravity \cite{Witten_2022,Chandrasekaran_20231,Chandrasekaran_20232,Kudler-Flam:2023qfl,Jensen:2023yxy}. In the context of entanglement wedge reconstruction it was shown that local modular flow implies an empty causal shadow -- the bulk causal wedge and the entanglement wedge must coincide \cite{Chen_2023,leutheusser2024subregionsubalgebradualityemergencespace}. 

On the other hand, non-local modular flows represent the generic scenarios. They are important for understanding novel features arising in the modular dynamics, which ultimately reflect the entanglement structure. For example they are essential for a causal shadow to be present, and thus underlie the capacity for the entanglement wedge reconstruction to go beyond bulk causality \cite{Chen_2023,leutheusser2024subregionsubalgebradualityemergencespace}. However, they are difficult to study and progresses in understanding them have been limited. The few well-understood cases include the modular flows for the multi-interval subregion in the vacuum of two-dimensional massless free fermion \cite{Casini2009Sep,Cardy:2016fqc,Arias_2018,Erdmenger:2020nop}in. In this case the non-locality is mild -- it mixes the local modular flows from the single-interval components. 
The more recent work \cite{Lashkari_2021} studied the modular Hamiltonians on the half-plane for a general class of states, whose non-local modular flows simplify for the squeezed states in the generalized free field theory due to the truncation of the Baker–Campbell–Hausdorff expansion.

Studying non-local modular flows in full generality is a daunting task. In this paper, we study a particular type of non-local modular flows that nonetheless represent a class with reasonable generality.
They are generated by modular Hamiltonians associated with the vacuum state, but across deformed null cuts: 
\be
\rho^\Omega_{\tilde{A}} = \exp{\left(-H^\Omega_{\tilde{A}}\right)},\qquad\partial\tilde{A} = \left\lbrace (x^+,x^-,x^{\bot})\in \mathbb{R}^{1,d-1} \,|\, x^-=0,\; x^+=\gamma(x^{\perp})\right\rbrace .
\ee
For simplicity, in this paper we focus on the case of free scalar QFTs in 2+1 dimensional space-time (our method is in fact valid for any dimension $d\ge 2$). This is the simplest kinematic setting that we hope can capture the main features associated with this type of non-local modular flows. 

The goal of this work is summarized as follows. Firstly, we hope to perform computations regarding non-local modular flows in as much explicit terms as possible. Secondly, in doing this we also hope to gain some qualitative understanding, and in particular to compare against existing pictures concerning generic modular flows -- for example, that when applied to operators localized near the entangling surface it can be approximated by local boosts appropriately defined. The hope is that our results can provide an opportunity to explicitly examine and clarify the approximations behind these existing intuitions. 

This paper is structured in the following way. In \textbf{section} \ref{sec:review} we review some basics about half-sided modular inclusion, a property that allows us to write down the relevant modular Hamiltonians in closed form. In \textbf{section} \ref{sec:generator} we carry out explicit computations of the commutators between modular Hamiltonians and local fields, which gives the generators of the modular flows. In \textbf{section} \ref{sec:flow} we apply the commutator results to study the finite modular flows. We conclude the paper with some discussions in \textbf{section} \ref{sec:discussion}, emphasizing on the relation between our results and existing intuitions about approximate modular flows near the entangling surface.   

\section{Modular Hamiltonian from half-sided modular inclusion}\label{sec:review}
There are several difficulties associated with studying general modular flows.
The very first one concerns with the lack of explicit forms for the modular Hamiltonian in the general cases.
However, significant progresses can be made in a special class of cases corresponding to the vacuum state across a deformed null-cut. In this section, we provide a very briefly review about this. 

\begin{figure}[h]
    \begin{center}
    \includegraphics[scale=0.8]{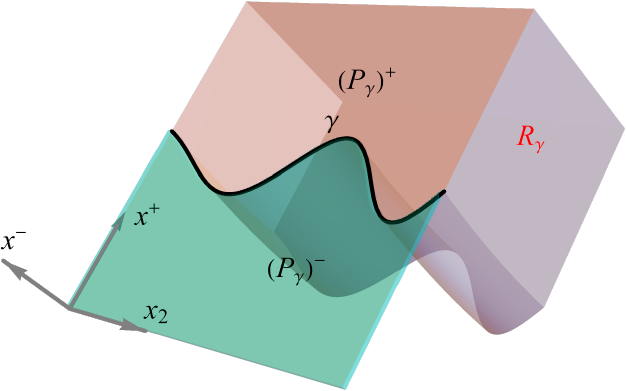}
    \end{center}
    \caption{The null plane is split into two parts $P_{\gamma}^+,\ P_{\gamma}^-$ by a null-cut $x^+=\gamma(x^\bot)$. The causal completion to the right hand side $(x^-<0)$ across $\gamma$ is denoted by $\mathcal{R}_{\gamma}$. }
    \label{fig: null-cut}
\end{figure}

To begin with, consider a quantum field theory on $\mathbb{R}^{1,d-1}$ with the standard metric $\eta_{\mu\nu} = \text{diag}(-1,1,...,1)$,
and $\ket{\Omega}$ is its ground state.
Take a null plane $\mathcal{P}:\ x^-=0\ (x^{\pm}:=x^0\pm x^1)$ and a codimension-2 curve $\gamma\subset\mathcal{P}$ on it: $x^+=\gamma(x^\bot)$.
Then $\gamma$ splits $\mathcal{P}$ into two parts:
the future part $\mathcal{P}_{\gamma}^+=\{x\in\mathcal{P}\ |\ x^+>\gamma(x^\bot)\}$ and the past part $\mathcal{P}_{\gamma}^-=\{x\in\mathcal{P}\ |\ x^+<\gamma(x^\bot)\}$.
We consider the causal completion of the future part $\mathcal{P}_{\gamma}^+$, i.e. $\mathcal{R}_{\gamma}:=\mathcal{P}_{\gamma}^{+\prime\prime} $,
where symbol $ ^{\prime}$ means the causal complement.
In this case, the modular Hamiltonian $H_{\mathcal{R}_{\gamma}}$ associated to the region $\mathcal{R}_{\gamma}$ and state $\ket{\Omega}$ is given by \cite{Casini_2017,HWang2019Sep},
\begin{equation}
H_{\mathcal{R}_{\gamma}} = 2\pi\int\dd^{d-2}x^{\bot}\int_{\gamma(x^{\bot})}^{+\infty}\dd x^+\ \left(x^+-\gamma(x^{\bot})\right)T_{++}(x^+,x^-=0,x^{\bot}).
\end{equation}
Due to the intrinsic UV sensitivity of the entanglement structures,
a more well-defined object in QFTs is given by the so-called full modular Hamiltonian,
defined as: 
\begin{equation}\label{eq: full modular Hamiltonian}
\begin{aligned}
    \hat{H}_{\gamma}&= H_{\mathcal{R}_{\gamma}} - H_{\mathcal{L}_{\gamma}} \\
    &= 2\pi\int\dd^{d-2}x^{\bot}\int_{-\infty}^{+\infty}\dd x^+\ \left(x^+-\gamma(x^{\bot})\right)T_{++}(x^+,x^-=0,x^{\bot}),
\end{aligned}
\end{equation}
where $\mathcal{L}_{\gamma}$ is the causal completion of the past part $\mathcal{P}_{\gamma}^{-}$: $\mathcal{L}_{\gamma}:=\mathcal{P}_{\gamma}^{-\prime\prime} $.
We summarize the above setup in the figure \ref{fig: null-cut}.

In the process of deriving the above expression \eqref{eq: full modular Hamiltonian},
a remarkable algebraic property knwon as the \emph{half-sided modular inclusion} played a crucial role \cite{Casini_2017,HWang2019Sep}.
It is well known that the full modular Hamiltonian for Rindler space is proportional to the boost operator $M_{10}$,
\begin{equation}
    \hat{H}_0=2\pi\int\dd^{d-2}x^{\bot}\int_{-\infty}^{+\infty}\dd x^+\ x^+T_{++}(x^+,x^-=0,x^{\bot}),
\end{equation}
and its action on the field operator $\phi(x)$ is given by
\begin{equation}
    e^{i\hat{H}_0t}\phi(x)e^{-i\hat{H}_0t}=\phi(e^{2\pi t}x^+,e^{-2\pi t}x^-,x^{\bot}).
\end{equation}
Without loss of generality, we assume the null-cut $\gamma>0$.
Then, the von Neumann algebra associated to the regions $\mathcal{R}_0,\ \mathcal{R}_{\gamma}$ form a \emph{half sided modular inclusion}
$(\mathcal{R}_0,\mathcal{R}_{\gamma},\ket{\Omega})$ \cite{borchers2000revolutionizing}, which is defined by the property: 
\begin{equation*}
    e^{is\hat{H}_0}\mathcal{R}_{\gamma}e^{-is\hat{H}_0}\subset \mathcal{R}_{\gamma},\quad \forall\; s\ge 0.
\end{equation*}
Correspondingly, there exists an unique \emph{half sided translation} $U(s)=\exp(-iP_{\gamma}s)$, which is a one-parameter of continuous unitary group satisfying: 
\begin{itemize}
    \item $P_{\gamma}$ is a semi-definite operator.
    \item $P_{\gamma}\ket{\Omega}=0$.
    \item $U^\dagger(s)\mathcal{R}_{0}U(s)\subset \mathcal{R}_{0},\quad \forall s\ge 0$.
    \item $\mathcal{R}_{\gamma}=U(1)^{\dagger}\mathcal{R}_0U(1)$
\end{itemize}
For CFTs or more generally QFTs obtained from relevant deformations of CFT fixed points,
\cite{Casini_2017} derived the commutators of all kinds of operators constructed from integrating the stress-energy tensor $T_{\mu\nu}$ along the null line $x^+$,
by utilizing an analysis similar to one use for the operator product expansion.
Subsequently, the algebraic properties of half-sided translation can uniquely determine the $P_{\gamma}$.
Besides, \cite{HWang2019Sep} chosed another approach to derive the $P_{\gamma}$, where we review here.
Since $(\mathcal{R}_0,\mathcal{R}_{\gamma},\ket{\Omega})$ is a half sided modular inclusion,
we have two algebraic relations,
\begin{align}
    e^{-i\hat{H}_0t}\hat{H}_{\gamma} e^{i\hat{H}_0t}=\hat{H}_{e^{2\pi t}\gamma},\\
    [\hat{H}_{0},\hat{H}_{\gamma}]=-2\pi i(\hat{H}_{0}-\hat{H}_{\gamma}).
\end{align}
The above two formulae can be gathered into an evolution equation for $\hat{H}_{\lambda \gamma}$ generated by boost operator $\hat{H}_0$,
\begin{equation}
    \lambda\frac{\dd}{\dd\lambda}(\hat{H}_{\lambda \gamma}-\hat{H}_0)=(\hat{H}_{\lambda \gamma}-\hat{H}_0).
\end{equation}
This equation can be solved as,
\begin{equation}
    \hat{H}_{\lambda \gamma}=\hat{H}_0-2\pi\lambda P_{\gamma}.
\end{equation}
Meanwhile, the perturbative results of \cite{HWang2016Sep} showed that to first order in the shape deformation $x^+=\gamma(x^{\bot})$ one has,
\begin{equation}\label{eq: perturbative H gamma}
    \hat{H}_{\gamma}=\hat{H}_0-2\pi  \int\dd^{d-2} x^{\bot}\int_{-\infty}^{+\infty}\dd x^+\ \gamma(x^{\bot})T_{++}(x^+,x^-=0,x^{\bot})+O(\gamma^2).
\end{equation}
Take $\lambda$ small we find that the perturbative result \eqref{eq: perturbative H gamma} truncates to the first order for null deformation,
and
\begin{equation}\label{eq: P gamma}
    P_{\gamma}=\int\dd^{d-2}x^{\bot}\int_{-\infty}^{+\infty}\dd x^+\ \gamma(x^{\bot})T_{++}(x^+,x^-=0,x^{\bot}).
\end{equation}
Eventually, we obtain the full modular Hamiltonian $\hat{H}_{\gamma}$ \eqref{eq: full modular Hamiltonian}.
We conclude this section by making some comments regarding the proof of the \eqref{eq: full modular Hamiltonian}. At the technical level it may appear that the derivation based on the half-sided modular inclusion of $(\mathcal{R}_0,\mathcal{R}_{\gamma},\ket{\Omega})$ requires the null-cut $\gamma(x^{\bot})$ to be positive definite, or at least bounded from below.  Additional arguments based on the entanglement structure on the null plane suggests that \eqref{eq: full modular Hamiltonian} remains valid for arbitrary null cut $\gamma(x^{\bot})$ \cite{Casini_2017,Casini:2018kzx}. In particular, it can be shown that for general relativistic QFTs, the vacuum entanglement entropies of any sub-regions $U$ and $V$ whose boundaries are on the null plane $x^-=0$ saturate the strong sub-additivity bound:
\begin{equation}
    S(U)+S(V)-S(U\cap V)-S(U\cup V)=0.
\end{equation}
The fact that this holds for arbitrary such $U$ and $V$ then implies that when viewed as a state on the null-plane, the vacuum is Markovian, i.e. there exists no entanglement between the null ray generators. As a result, the reduced density matrix can be heuristically thought of as a tensor product over the null rays. The corresponding modular Hamiltonian is therefore an additive sum over contributions from these null rays, i.e. of the form (\ref{eq: full modular Hamiltonian}) for any shape of the null-cut $\gamma$. 

\section{Modular commutator as modular-flow generator}\label{sec:generator}
In this section, we firstly compute the commutator $[\hat{H}_{\gamma},\phi(x)]$ in free scalar theory for arbitrary null-cuts $\gamma$,
employing the mode expansion method.
Subsequently, we consider specific forms of $\gamma$ to perform further calculations.
Throughout this process,
we obtain some examples of the commutator $[\hat{H},\phi(x)]$ and identify conditions under which the modular flows become local or non-local.
Finally, we extend our findings to null-cuts expressed as power functions.

\subsection{Mode expansion of the modular commutator}
Suppose $\phi(x)$ is a free massive scalar field in $3$-dimensional Minkowski spacetime $\mathbb{R}^{1,2}$,
\begin{equation}
    \mathcal{L} = -\frac{1}{2}\left(\partial\phi\right)^2 - \frac{1}{2}m^2\phi^2,
\end{equation}
where we take dimension equals to $3$ just for simplicity.
In fact, our computation holds for any dimension $d> 2$.
Then we adopt a general codimension-2 curve $x^+=\gamma(x^{\bot})$ on the null plane $\mathcal{P}:\ x^-=0$,
which splits the null plane into two subregions $\mathcal{P}_{\gamma}^+$ and $\mathcal{P}_{\gamma}^-$.
Currently, the full modular Hamiltonian $\hat{H}_{\gamma}$ associated to the partition is given by \eqref{eq: full modular Hamiltonian}.
For the field operator $\phi(x)$,
its modular evolution is given by Baker–Campbell–Hausdorff formula,
\begin{equation}
    \phi(s,x) = e^{i\hat{H}_{\gamma}s}\phi(x)e^{-i\hat{H}_{\gamma}s}=e^{is\,\mathrm{ad}\hat{H}_{\gamma}}\phi(x).
\end{equation}
Consequently, the first step towards understanding the modular flow is to calculate the modular commutator,
\begin{equation}
    \partial_s\phi(s,x) = i[\hat{H}_{\gamma},\phi(s,x)].
\end{equation}
Since $\hat{H}_{\gamma}= M_{10}-2\pi P_{\gamma}$,
we only need to compute $[P_{\gamma},\phi(x)]$.

In comparison with field operators,
it is more convenient to deal with correlation functions.
Therefore, we rewrite the commutator $[P_{\gamma},\phi(x)]$ in terms of correlation functions.
For free theory, field operator $\phi(x)$ and its conjugate momentum $\pi(x)=\partial_0\phi(x)$ form a complete canonical basis on a Cauchy slice,
\begin{equation}
    [P_{\gamma},\phi(x^0,\vec{x})]=\int_{y^0=x^0} \dd^{d-1} y \ D_1(x,y)\phi(y)+D_2(x,y)\pi(y),
\end{equation}
where kernel function $D_1(x,y)$ and $D_2(x,y)$ can be determined by the canonical commutator relation $[\phi(x^0,\vec{x}),\pi(x^0,\vec{y})]=i\delta^{(d-1)}(\vec{x}-\vec{y})$:
\begin{align}
    D_1(x,y)|_{y^0=x^0}&=-i\left[[P_{\gamma},\phi(x^0,\vec{x})],\pi(x^0,\vec{y})\right],\\
    D_2(x,y)|_{y^0=x^0}&=i\left[[P_{\gamma},\phi(x^0,\vec{x})],\phi(x^0,\vec{y})\right].
\end{align}
Since $D_1(x,y)$ and $D_2(x,y)$ are $\mathbb{C}-$numbers,
we can put them into the vacuum expectation value,
\begin{equation}\label{eq:Definitin A}
    \begin{aligned}
    \left[[P_{\gamma},\phi(x^0,\vec{x})],\phi(x^0,\vec{y})\right]&=-\braket{0|\phi(x^0,\vec{x})P_{ \gamma}\phi(x^0,\vec{y})|0}-\braket{0|\phi(x^0,\vec{y})P_{ \gamma}\phi(x^0,\vec{x})|0}\\
    &=:-A-A^*.
    \end{aligned}
\end{equation}
Similarly,
\begin{equation}\label{eq:Definitin B}
    \begin{aligned}
        \left[[P_{\gamma},\phi(x^0,\vec{x})],\pi(x^0,\vec{y})\right]&=-\braket{0|\phi(x^0,\vec{x})P_{ \gamma}\pi(x^0,\vec{y})|0}-\braket{0|\pi(x^0,\vec{y})P_{ \gamma}\phi(x^0,\vec{x})|0}\\
            &=:-B-B^*.
    \end{aligned}
\end{equation}
    To sum up,
\begin{equation}\label{eq:commutator to correlator}
    [P_{\gamma},\phi(x)]=i\int_{y^0=x^0} \dd^{d-1} y\ \left[(B+B^*)\phi(x^0,\vec{y})-(A+A^*)\pi(x^0,\vec{y})\right].
\end{equation}

Before we start the calculations of $A,B$,
there is a subtlety need to be addressed.
We know the energy momentum tensor $T_{\mu\nu}$ derived from N\"other theorem is given by,
\begin{equation}
    T_{\mu\nu} = \partial_{\mu}\phi\partial_{\nu}\phi-\frac{1}{2}\eta_{\mu\nu}\left(\partial\phi\right)^2-\frac{1}{2}\eta_{\mu\nu}m^2\phi^2.
\end{equation}
However, when $m=0$, $T_{\mu\nu}$ is not traceless,
meaning it fails to describe the conformal field theory (free massless scalar field theory).
In fact, the traceless energy momentum tensor $\tilde{T}_{\mu\nu}$ for $m=0$ is,
\begin{equation}
    \tilde{T}_{\mu\nu} = \partial_{\mu}\phi\partial_{\nu}\phi-\frac{1}{2}\eta_{\mu\nu}\left(\partial\phi\right)^2-\frac{1}{8}(\partial_{\mu}\partial_{\nu}-\eta_{\mu\nu}\partial^2)\phi^2.
\end{equation}
Note that in the definition of $P_{\gamma}$ \eqref{eq: P gamma},
only the $T_{++}$ or $\tilde{T}_{++}$ is involved,
which respectively are,
\begin{align}
    T_{++}&=\partial_+\phi\partial_+\phi,\\
    \tilde{T}_{++}&=\partial_+\phi\partial_+\phi-\frac{1}{4}\partial_+(\phi\partial_+\phi).
\end{align}
It is straightforwardly to see that $\partial_+(\phi\partial_+\phi)$ will produce a vanishing boundary term when substituting $\tilde{T}_{++}$ into $P_{\gamma}$.
Therefore, $T_{++}$ and $\tilde{T}_{++}$ will yield the same $P_{\gamma}$ for $m=0$.
We can securely use
\begin{equation}\label{eq: P gamma 2}
    P_{\gamma} = \int\dd z^+\dd z_2\ \gamma(z_2):\partial_+\phi\partial_+\phi:(\frac{z^+}{2},\frac{z^+}{2},z_2)
\end{equation}
in the following analysis.

The free theory is solvable,
which is reflected in the fact that we can expand $\phi(x)$ in terms of creation and annihilation operators,
\begin{equation}
    \phi(x) = \int\frac{\dd^2p}{(2\pi)^2}\frac{1}{\sqrt{2E_{\vec{p}}}}\left(a_{\vec{p}}\,e^{ip\cdot x} + a^{\dagger}_{\vec{p}}\,e^{-ip\cdot x}\right),\;\; E_{\vec{p}} = \sqrt{m^2+|\vec{p}|^2}
\end{equation}
After substituting above equations into \eqref{eq: P gamma 2}, we directly obtain\footnote{For the convenience of presentation,
we shall not distinguish between the covariant (upper) and contravariant (lower) indices for the spatial components $x^i=x_i,\,1\le i\le d-1$.}
\begin{multline}
    P_{\gamma}=-\frac{1}{4}\int\dd z^+\dd z_2\ \gamma(z_2)\int\frac{\dd^2 p\dd^2 q}{(2\pi)^4}\frac{p^-q^-}{\sqrt{2E_{\vec{p}}2E_{\vec{q}}}} \times\Big[a_{\vec{p}}\,a_{\vec{q}}\,e^{-i\frac{(p^-+q^-)z^+}{2}+i(p_2+q_2)z_2}\\
        -a^\dagger_{\vec{q}}\,a_{\vec{p}}\,e^{{-i\frac{(p^--q^-)z^+}{2}+i(p_2-q_2)z_2}}-a^\dagger_{\vec{p}}\,a_{\vec{q}}\,e^{{i\frac{(p^--q^-)z^+}{2}-i(p_2-q_2)z_2}}+a^\dagger_{\vec{p}}\,a^\dagger_{\vec{q}}\,e^{{i\frac{(p^-+q^-)z^+}{2}-i(p_2+q_2)z_2}}\Big],
\end{multline}
where $p^{\pm}:=p^0\pm p^1 = E_{\vec{p}}\pm p^1$.
If we use the Fourier transformation to expand the codimension-2 curve $x^+=\gamma(z_2)$,
the above equation will become intricate.
However, it is noteworthy that whenever $\gamma(z_2)$ is combined with a plane wave $e^{ip\cdot z}$,
it can be transformed into a differential operator,
\begin{equation}
    \gamma(z_2)e^{ip_2z_2}=\gamma\left(\frac{1}{i}\frac{\partial}{\partial p_2}\right)e^{ip_2z_2}.
\end{equation}
The mode expansion of $P_{\gamma}$ is now observed to resemble a diagonalized quadratic of creation and annihilation operators,
\begin{equation}
    \begin{aligned}
        P_{\gamma}&=-\frac{1}{2}\int\frac{\dd^2 p\dd^2 q}{(2\pi)^2}\frac{p^-q^-}{\sqrt{2E_{\vec{p}}2E_{\vec{q}}}}\\
        &\times \Big[a_{\vec{p}}\,a_{\vec{q}}\,\delta(p^-+q^-)\gamma\left(\frac{1}{i}\frac{\partial}{\partial p_2}\right)\delta(p_2+q_2)-a^\dagger_{\vec{q}}\,a_{\vec{p}}\,\delta(p^--q^-)\gamma\left(\frac{1}{i}\frac{\partial}{\partial p_2}\right)\delta(p_2-q_2)\\
        &-a^\dagger_{\vec{p}}\,a_{\vec{q}}\,\delta(p^--q^-)\gamma\left(\frac{-1}{i}\frac{\partial}{\partial p_2}\right)\delta(p_2-q_2)+a^\dagger_{\vec{p}}\,a^\dagger_{\vec{q}}\,\delta(p^-+q^-)\gamma\left(\frac{-1}{i}\frac{\partial}{\partial p_2}\right)\delta(p_2+q_2)\Big].
    \end{aligned}
\end{equation}
Subsequently, we can compute $A$ in \eqref{eq:Definitin A},
\begin{equation}
    \begin{aligned}
        A&=\Braket{0|\phi(x^0-i\epsilon,\vec{x}) P_{\gamma}\phi(y^0+i\epsilon,\vec{y})|0 }\\
        &=\frac{1}{8}\int\frac{\dd^2 p\dd^2 q}{(2\pi)^2}\frac{p^-q^-}{E_{\vec{p}}E_{\vec{q}}}\delta(p^--q^-)\\
        &\times\Big[e^{iq\cdot x}e^{-ip\cdot y}\gamma\left(\frac{1}{i}\frac{\partial}{\partial p_2}\right)\delta(p_2-q_2)
        +e^{ip\cdot x}e^{-iq\cdot y}\gamma\left(\frac{-1}{i}\frac{\partial}{\partial p_2}\right)\delta(p_2-q_2)\Big],
    \end{aligned}
\end{equation}
where we use the $i\epsilon$ prescription to make sure the integration by parts valid in later calculations.
Unlike equation \eqref{eq:Definitin A}, here we reserve $x^0$ and $y^0$ as two independent variables, which will be needed for the computation of $B$ in \eqref{eq:Definitin B}.
We firstly integrate $q_1$ out,
by regarding $\delta(q^--p^-)$ as a $\delta-$function with argument $q_1$,
\begin{equation}
    \delta(q^--p^-)=\frac{E_{\vec{q}}}{q^-}\delta(q_1-q_1^*),
\end{equation}
where
\begin{equation}\label{eq: q1*}
    q_1^*=\frac{p_1\left(2m^2+(p_2)^2+(q_2)^2\right)+\sqrt{m^2+(p_1)^2+(p_2)^2}\left((q_2)^2-(p_2)^2\right)}{2(m^2+(p_2)^2)}.
\end{equation}
We have,
\begin{align}\label{eq:A}
    A=\frac{1}{8}\int&\frac{\dd^2 p\dd q_2}{(2\pi)^2}\frac{p^-}{E_{\vec{p}}}\times\Big[e^{-i\sqrt{(q_1^*)^2+(q_2)^2+m^2}\cdot (x^0-i\epsilon)+iq_1^*\cdot x_1+iq_2\cdot x_2}e^{-ip\cdot y}{\gamma\left(\frac{1}{i}\frac{\partial}{\partial p_2}\right)\delta(p_2-q_2)}\notag\\
    &+e^{ip\cdot x}e^{i\sqrt{(q_1^*)^2+(q_2)^2+m^2}\cdot (y^0+i\epsilon)-iq_1^*\cdot y_1+iq_2\cdot y_2}{\gamma\left(\frac{-1}{i}\frac{\partial}{\partial p_2}\right)\delta(p_2-q_2)}\Big].
\end{align}
The next step is to substitute the partial derivative with respect to $p_2$ with a partial derivative with respect to $q_2$,
\begin{equation}
    \left(\frac{1}{i}\frac{\partial}{\partial p_2}\right)\delta(p_2-q_2)=\left(-\frac{1}{i}\frac{\partial}{\partial q_2}\right)\delta(p_2-q_2).
\end{equation}
Then we can perform integration by parts on $q_2$,
which is easier than directly integrating by parts on $p_2$.
\begin{align}\label{eq:A}
    A=\frac{1}{8}\int\frac{\dd^2 p}{(2\pi)^2}\frac{p^-}{E_{\vec{p}}}&\times\Big[{\gamma\left(\frac{1}{i}\frac{\partial}{\partial q_2}\right)\Big|_{q_2=p_2}}e^{-i\sqrt{({q_1^*})^2+({q_2})^2+m^2}\cdot (x^0-i\epsilon)+i{q_1^*}\cdot x_1+i{q_2}\cdot x_2}e^{-ip\cdot y}\notag\\
        &+{\gamma\left(\frac{-1}{i}\frac{\partial}{\partial q_2}\right)\Big|_{q_2=p_2}}e^{i p\cdot x}e^{i\sqrt{({q_1^*})^2+({q_2})^2+m^2}\cdot (y^0+i\epsilon)-i{q_1^*}\cdot y_1-i{q_2}\cdot y_2}\Big].
\end{align}
Please note that the derivative operator $\gamma(\frac{\partial}{\partial q_2})$ should act on $q_1^*,q_2$,
and then set $q_2=p_2$.
After deriving the expression for $A$,
we can obtain $B$ \eqref{eq:Definitin B} by taking the derivative on $A$ with respect to $y^0$,
\begin{equation}\label{eq:B}
    \begin{aligned}
        B&= \Braket{0|\phi(x^0-i\epsilon,\vec{x}) P_{\gamma}\partial_{y^0}\phi(y^0+i\epsilon,\vec{y})|0 }\\
        &=\frac{i}{8}\int\frac{\dd^2 p}{(2\pi)^2}\frac{p^-}{E_{\vec{p}}}\times\Big[E_{\vec{p}}\,{\gamma\left(\frac{1}{i}\frac{\partial}{\partial q_2}\right)\Big|_{q_2=p_2}}e^{-i\sqrt{({q_1^*})^2+({q_2})^2+m^2}\cdot (x^0-i\epsilon)+i{q_1^*}\cdot x_1+i{q_2}\cdot x_2}e^{-ip\cdot y}\\
        &+{\gamma\left(\frac{-1}{i}\frac{\partial}{\partial q_2}\right)\Big|_{q_2=p_2}}\sqrt{({q_1^*})^2+({q_2})^2+m^2}e^{i p\cdot x}e^{i\sqrt{({q_1^*})^2+({q_2})^2+m^2}\cdot (y^0+i\epsilon)-i{q_1^*}\cdot y_1-i{q_2}\cdot y_2}\Big].
    \end{aligned}
\end{equation}
So far, we have derived the expressions of $A$ and $B$ in momentum space.
Given the null-cut $x^+=\gamma(x_2)$,
we can complete integrals of $\vec{p}$ in \eqref{eq:A} and \eqref{eq:B},
and then substitute the results of $A$ and $B$ into \eqref{eq:commutator to correlator} to obtain the final expression of $[P_{\gamma},\phi(x)]$.
Combined with the Rindler boost $2\pi M_{10}$,
we can determine the infinitesimal modular flow $[\hat{H}_{\gamma},\phi(x)]=2\pi[M_{10}-P_{\gamma},\phi(x)]$.

\subsection{Examples: monomial profiles $\gamma(x)=x^n$}
We now proceed to compute the commutator $[P_{\gamma},\phi(x)]$ explicitly by choosing specific forms for the null-cuts $\gamma(x_2)$. To be concrete,
we focus on the monomial function $\gamma(x)=x^n$.
The computation is lengthy but straightforward, we will provide some of the details in the \textbf{Appendix} \ref{sec:appendix z2square}.
One observation about the computation worth remarking is that although the individual terms of $A, A^*, B, B^*$ may be complex,
the final results only depend on the combinations $A + A^*$ and $B + B^*$, which simplifies significantly. For illustration, we list below the results for $\gamma(x_2)\in\left\{ 1,\,x_2,\,(x_2)^2,\,(x_2)^3\right\}$. 

\begin{enumerate}
    \item [(i).] $\gamma(x_2)=1$:
    \begin{equation}\label{eq:gamma=1}
        [P_{\gamma},\phi(x)]=[P_+,\phi(x)]=-i\partial_+\phi(x).
    \end{equation}
    \item [(ii).] $\gamma(x_2)=x_2$:
    \begin{equation}\label{eq:gamma=z2}
        [P_{\gamma},\phi(x)]=-[M_{+2},\phi(x)]=i\left(x_+\partial_2-x_2\partial_+\right)\phi(x).
    \end{equation}
    \item [(iii).] $\gamma(x_2)=(x_2)^2$:   
    \begin{equation}\label{eq:gamma=z2square}
        \begin{aligned}
            [P_{\gamma},\phi(x)]&=i\left(-|x|^2\partial_++2x_+(x\cdot\partial)+x_+\right)\phi(x)\\
            &+\frac{i}{2} m (x^-)^2\partial_{x^-}\int\dd y_2\ e^{-m|x_2-y_2|}\phi(x^0,x_1,y_2).
        \end{aligned}
    \end{equation}
     \item [(iv).] $\gamma(x_2)=(x_2)^3,\ m=0$:
    \begin{align}\label{eq:gamma=z2cube}
        [P_{\gamma},\phi(x)]&|_{m=0}=i\left[-(x_2)^3\partial_{+}-3(x^-)^2x_2\partial_{-}-\frac{1}{2}x^-((x^-)^2+3(x_2)^2)\partial_{2}-\frac{3}{2}x^-x_2\right]\phi(x)\notag\\
        &+i(x^-)^2\left(-\frac{3}{2}+x^-\partial_{x_1}\right)\partial_{x^-}\int\dd y_2\ \mathrm{sgn}(x_2-y_2)\phi(x^0,x_1,y_2).
    \end{align}
\end{enumerate}
Here for simpler presentation we have chosen to only show the $m=0$ result for the cubic profile $\gamma(x_2) =(x_2)^3$ case. Despite the apparent dependence on only the field operators $\phi(x)$, the right-hand-sides have also included the contributions from the conjugate momentum operators $\Pi(x)$ by the derivatives $\partial_{\pm}$. 

We can make a few observations. Firstly, the modular flow is local for the cases of $\gamma(x_2)\in\left\{ 1,\,x_2\right\}$; it is also local for $\gamma(x_2)=(x_2)^2$ at $m=0$ -- the non-local part is proportional to $m$. These local modular flows share a common origin that is related to the symmetry of the set-up. To see this, we label the cases of $\gamma(x_2)\in \{1,\,x_2,\,(x_2)^2\}$ by $\{P_1,\,P_{x_2},\,P_{(x_2)^2}\}$ respectively. Then the half-sided translations can be easily identified as the following charges written on the null surface,
\be
P_1 = \hat{P}_+,\qquad\quad P_{x_2}=\hat{M}_{2+},\qquad\quad P_{(x_2)^2}=\hat{K}_+,
\ee
where $\hat{P}_+$ is the generator of the null translation; $\hat{M}_{2+}$ is the generator of boost along $(+,2)$ direction; and finally $\hat{K}_+$ is the generator of the special conformal transformation. The first two operators $P_1,\ P_{x_2}$ are conserved charges for any relativistic QFTs, and they generate isometries; the operator $P_{x^2}$ is a conserved charge if the underlying theory is a CFT, and it generates conformal isometries. They are members of the conformal symmetries whose conformal killing vectors are, 
\begin{equation}
    \begin{aligned}
        &p_{\mu}=-i\partial_{\mu},\qquad d_{\mu}=-x\cdot\partial,\\
        &m_{\mu\nu}=-i(x_{\mu}\partial_{\nu}-x_{\nu}\partial_{\mu}),\qquad k_{\mu}=i\left(2x_{\mu}(x\cdot\partial)-|x|^2\partial_{\mu}\right).
    \end{aligned}
\end{equation}
When acting on a primary scalar field $\phi(x)$ with scaling dimension $\Delta$ in a CFT,
the corresponding charges take the following commutation relations:
\begin{equation}
    \begin{aligned}
        [\hat{P}_{\mu},\phi(x)]&=-i\partial_{\mu}\phi(x),\\
        [\hat{D},\phi(x)]&=(-x\cdot\partial+\Delta)\phi(x),\\
        [\hat{M}_{\mu\nu},\phi(x)]&=-i(x_{\mu}\partial_{\nu}-x_{\nu}\partial_{\mu})\phi(x),\\
        [\hat{K}_{\mu},\phi(x)]&=i\left(2x_{\mu}(x\cdot\partial)-|x|^2\partial_{\mu}+2\Delta x_{\mu}\right)\phi(x).
    \end{aligned}
\end{equation}
The RHS of (\ref{eq:gamma=1}, \ref{eq:gamma=z2}), and the $m=0$ limit (thus a CFT) of (\ref{eq:gamma=z2square}) agrees precisely with the corresponding actions of the conserved charges. 

While the results showing local modular flows serve as a consistency check for our computations, they also manifest the more general conclusion \cite{Sorce_2024,Chen_2023}. It states that for a (conformal) QFT, the modular flow of $K^\Omega_A$ associated with the state $\Omega$ and subsystem $A$ (with causal completion $\mathcal{D}$) is local only if both $\mathcal{D}$ and $\Omega$ are invariant under a class of (conformal) isometries. The local modular flows then simply follow the integral curves of the (conformal) killing vector fields. One can also interpret these as the facts that for the null cuts $\gamma(x_2)\in\{1,\,x_2\}$, the set-ups can be mapped to the vacuum half-plane (whose modular flows are local boosts) by a Poincar\'e transformation; while for $\gamma(x_2)=(x_2)^2$ it can be mapped to the vacuum half-plane by a special conformal transformation if the theory is conformal ($m=0$). This also explains why the modular flows associated with $\gamma(x_2)=(x_2)^n,\;n\geq 3$ are non-local even for the massless case.

A second observation is that the non-local terms, e.g. those in $\gamma(x_2)=(x_2)^2$ at $m\neq 0$ and $\gamma(x_2)=(x_2)^3$, are all proportional to powers of $x^-$. As a result, the non-local effects diminish as powers of the transverse distance $x^-$ to the entangling null surface. This agrees with the general expectation. When restricted on the null plane $x^-=0$ the modular flow becomes local in general, and the local terms simplify to: 
\begin{equation}
    [P_{(x_2)^n},\phi]=-i (x_2)^n \partial_+ \phi\ \Longrightarrow\ [P_{\gamma},\phi] =-i \gamma(x_2) \partial_+ \phi .
\end{equation}
When combined to form the full modular Hamiltonian, the total modular flow on the null-plane takes the form: 
\begin{equation}
[\hat{H}_\gamma,\phi] = -2\pi i\left(x^+-\gamma(x_2)\right) \partial_+ \phi(x).
\end{equation}
These are local boosts along independent null rays $x^-=0$ labeled by constant $x_2$, and they starts from the entangling surface $x^+ = \gamma(x_2)$. This behavior is fully consistent with \cite{Casini_2017}, and reproducing it from our results provides another consistency check.  

\subsection{General non-local structure}
Based on the above examples,
we observe the general structure of the modular commutator $[\hat{H}_{\gamma},\phi(x)]$ as follows,
\begin{equation}\label{eq: general non-local}
    [\hat{H}_{\gamma}, \phi(x)] = \mathcal{A}(x, \partial_{x})\phi(x) + \sum_l (x^-)\mathcal{B}_l(x, \partial_{x})\int\dd y_2\ K_l(x_2, y_2)\phi(x^0, x_1, y_2),
\end{equation}
where $\mathcal{A}(x, \partial_{x})$ and $\mathcal{B}(x, \partial_{x})$ are differential operators, and $K(x_2 ,y_2)$ is a kernel function. We have also included a sum over $l$ to allow multiple terms of similar nature in the general cases. 
The main feature of (\ref{eq: general non-local}) is that the non-locality only emerges along the direction of $x_2$ in terms of a integration against the kernel function $K(x_2,y_2)$. In this section we provide some arguments for the general form of (\ref{eq: general non-local}) with respect to more general null cuts $\gamma(x_2)$. To begin with,
we note that $ [\hat{H}_{\gamma}, \phi(x)]$ is local for $\phi(x)$ restricted on the null plane $x^- = 0$,
which implies that the non-local terms must be proportional to $x^-$.
This justifies the factor of $x^-$ in the non-local part of \eqref{eq: general non-local}.
For simplicity, 
we will work explicitly on the slice $x^0=0$.

Based on the steps of computation discussed previously, we first analyze the term $A$ \eqref{eq:A}, where $\gamma(\partial/i\partial q_2)$ is understood by its Taylor expansion,
\begin{equation}  \label{eq:Taylor}
    \gamma\left(\frac{\partial}{\partial q_2}\right) = \sum_{n=0}^{\infty} \frac{\gamma^{(n)}(0)}{n!} \frac{\partial^n}{\partial q_2^n}.  
\end{equation}  
The basic ingredient of computing $A$ concerns how $\gamma(\partial/i\partial q_2)$ acts on $\exp( iq_1^*x_1+ iq_2x_2)$, or how $\gamma(-\partial/i\partial q_2)$ acts on $\exp(-iq_1^*y_1-iq_2y_2)$. Via its Taylor expansion (\ref{eq:Taylor}) we can focus on the monomials ${\partial^n}/{\partial q_2^n}$. Notably, the momentum $q_1^*$ in the exponent of is itself a quadratic polynomial of $q_2$ by \eqref{eq: q1*}. Consequently, after applying the differential operator ${\partial^n}/{\partial q_2^n}$, we obtain terms that are products of ${\partial q_1^*}/{\partial q_2}$, ${\partial^2 q_1^*}/{\partial (q_2)^2}$, $x_{1,2}$, and $y_{1,2}$. Working things out in more details, one can check that $A$ can be expressed as a linear combination of the following types of terms,
\begin{equation}\label{eq:A components}
\begin{aligned}
&e^{i\vec{p}\cdot(\vec{x} - \vec{y})}  \frac{p^-}{E_{\vec{p}}}\left(\frac{\partial q_1^*}{\partial q_2}\right)^{\alpha_1}\Big| _{q_2=p_2}\left(\frac{1}{i}\frac{\partial^2 q_1^*}{\partial (q_2)^2}\right)^{\alpha_2}\Big| _{q_2=p_2}(x_1)^{\alpha_1+\alpha_2}(x_2)^{n - \alpha_1 - 2\alpha_2},\\
\text{or}\qquad &e^{i\vec{p}\cdot(\vec{x} - \vec{y})}  \frac{p^-}{E_{\vec{p}}}\left(\frac{\partial q_1^*}{\partial q_2}\right)^{\alpha_1}\Big| _{q_2=p_2}\left(\frac{1}{i}\frac{\partial^2 q_1^*}{\partial (q_2)^2}\right)^{\alpha_2}\Big| _{q_2=p_2}(y_1)^{\alpha_1+\alpha_2}(y_2)^{n - \alpha_1 - 2\alpha_2},
\end{aligned}
\end{equation}
where $\alpha_1$ and $\alpha_2$ belong to $\mathbb{Z}_{\ge 0}$.
To simplify notation we omit for the moment writing the powers of  $x_1,x_2$ or $y_1,y_2$,
and focus on the common factors involving the momenta. We can then proceed the computation as follows: 
\begin{equation}
    \begin{aligned}
    \eqref{eq:A components} &= e^{i\vec{p}\cdot(\vec{x} - \vec{y})}\frac{(-i)^{\alpha_2}(p^+)^{\alpha_1 + \alpha_2 - 1}(p_2)^{\alpha_1}}{E_{\vec{p}}\left(m^2 + (p_2)^2\right)^{\alpha_1 + \alpha_2 - 1}}\\
    &= e^{i\vec{p}\cdot(\vec{x} - \vec{y})}\frac{(-i)^{\alpha_2}(p_2)^{\alpha_1}\left[\sum_{j=0}^{\alpha_1 + \alpha_2 - 1}\left( \begin{array}{c}
    \alpha_1 + \alpha_2 - 1 \\
    j \\
    \end{array} \right) E_{\vec{p}}^j(p_1)^{\alpha_1 + \alpha_2 - 1 - j}\right]}{E_{\vec{p}}\left(m^2 + (p_2)^2\right)^{\alpha_1 + \alpha_2 - 1}}.
    \end{aligned}
\end{equation}
where we remind that $p^+ = E_{\vec{p}} + p_1$. Suppose we replace $p_{1,2}$ in the numerator with $\partial/i\partial x_{1,2}$, then only terms containing an even number of $i$ will contribute to real part of $A$, which will be justified in \textbf{Appendix} \ref{sec:appendix z2square}. In other words, to obtain $A+A^*$, we only need to keep the terms in the summation over $j$ that involve an even number of occurrences of $i,p_1,p_2$. This requires $(2\alpha_1 + 2\alpha_2 - 1 - j)$ to be even,
i.e. $j$ is an odd integer. Because of this, we can set $j = 2\beta + 1$ for which $\beta \in \mathbb{Z}$, $0 \leq \beta \leq \frac{\alpha_1 + \alpha_2 - 2}{2}$.
It gives us the terms like
\begin{equation}
    \int\dd^2 p\ e^{i\vec{p}\cdot(\vec{x} - \vec{y})}\frac{\left((p_1)^2 + (p_2)^2 + m^2\right)^{\beta}i^{\alpha_2}(p_1)^{\alpha_1 + \alpha_2 - 2\beta - 2}(p_2)^{\alpha_1}}{\left(m^2 + (p_2)^2\right)^{\alpha_1 + \alpha_2 - 1}}.
\end{equation}
Since the exponent $\beta$ is an integer, we can proceed the computation by expanding the binomials in the numerator: 
\begin{equation}
 \left((p_1)^2+(p_2)^2+m^2\right)^\beta = \sum_{k=0}^{\beta}\left( \begin{array}{c}
    \beta \\
    k \\
    \end{array} \right) \left(m^2+(p_2)^2\right)^k(p_1)^{2(\beta-k)}  .
\end{equation}
For each of the resulting term the integrals with respect to $p_1$ and $p_2$ can be separated and performed independently. We can discuss two scenarios. The first possibility is that the powers of $m^2+(p_2)^2$ match between the numerator and denominator and thus cancel. This would produce a local term after the momentum integrals:
\begin{equation}\label{eq:partial_local}
    \int\dd^2 p\ e^{i\vec{p}\cdot(\vec{x} - \vec{y})} (p_1)^{\gamma_1} (p_2)^{\alpha_2}\propto \frac{\partial^{\gamma_1}}{\partial x_1^{\gamma_1}}\frac{\partial^{\alpha_2}}{\partial x_2^{\alpha_2}}\delta^{(2)}(\vec{x}-\vec{y}),
\end{equation}
where $\gamma_1,\gamma_2\in\mathbb{N}_{\ge0}$.
The other possibility is that a positive power of $m^2+(p_2)^2$ remains in the denominator -- recall that $\beta\leq (\alpha_1+\alpha_2)/2-1$. 
This will produce a non-local convolution whose kernel function depends only on $x_2 - y_2$:
\begin{eqnarray}\label{eq:partial_kernel}
 &&\int\dd^2 p\ e^{i\vec{p}\cdot(\vec{x} - \vec{y})}\frac{ (p_1)^{\gamma_3} (p_2)^{\alpha_2}}{((p_2)^2+m^2)^{\gamma_4}}\propto \frac{\partial^{\gamma_3}}{\partial x_1^{\gamma_3}}\delta(x_1-y_1) F(x_2-y_2),\nonumber\\
&& F(t) =\partial^{\alpha_2}_t\left(\int\dd p_2\ \frac{e^{i p_2 t}}{((p_2)^2+m^2)^{\gamma_4}}\right),
\end{eqnarray}
where $\gamma_3,\gamma_4\in\mathbb{N}_{\ge0}$. 

We now discuss the effects of putting back to powers of $x_{1,2}$ and $y_{1,2}$ in (\ref{eq:A components}). Regarding the powers of $x_1$ and $y_1$, we observe that in both (\ref{eq:partial_local}) and (\ref{eq:partial_kernel}) the term is  proportional to a finite order derivative of $\delta(x_1-y_1)$. This produces a local term along the $x_1$ direction if the order of the derivative is truncated, which according to (\ref{eq:Taylor}) is the case if we restrict to $\gamma(x_2)$ that are finite polynomials. As a result, the powers of $x_1=y_1$ enter the definitions of functions $\mathcal{A}$ and $\mathcal{B}$ in (\ref{eq: general non-local}). It is possible that non-locality along $x_1$ can arise from an infinite re-summation of (\ref{eq:Taylor}) for more general $\gamma(x_2)$, we leave this for the future investigations. Regarding the powers of $x_2$ and $y_2$, in the local case (\ref{eq:partial_local}) they enter the definition of the function $\mathcal{B}$ in (\ref{eq: general non-local}); in the non-local case (\ref{eq:partial_kernel}) they will break the translation invariance of $F(x_2-y_2)$ and make the kernel function a general two-variable function $K(x_2,y_2)$ \footnote{It happens that for sufficiently low orders of the polynomial $\gamma(x)$, the translation-invariance breaking does not occur, for example in the explicit results for $\gamma(x)=x^2$ and $\gamma(x)=x^3$.}. These are therefore consistent with the general form of (\ref{eq: general non-local}).
As for the computation of $B$ in  \eqref{eq:B},
we observe that:
\begin{equation}
    \frac{\partial ^l}{\partial ( q_2 ) ^l}\Big|_{q_2=p_2}\sqrt{( q_{1}^{*} ) ^2 + ( q_2) ^2 + m^2} = \frac{\partial ^l}{\partial ( q_2 ) ^l}\Big|_{q_2=p_2}q_1^*.
\end{equation}
Then the subsequent analysis is similar to that of $A$, which we will not repeat in details here. We therefore conclude that the general form of (\ref{eq: general non-local}) is valid for more general $\gamma(x_2)$ -- at least those of the form of finite polynomials.

We end this section by pointing out an interesting subtlety related to the non-local term appearing in the modular commutator. 
In the context of the algebraic QFTs, the Tomita-Takesaki theory \cite{borchers2000revolutionizing} dictates that the von-Neumann algebra is invariant under the corresponding modular flow: 
\be
 e^{i\hat{H}_Us} \mathcal{O} e^{-i\hat{H}_Us} \in \mathcal{A}_U,\qquad \forall\mathcal{O}\in \mathcal{A}_U ,
\ee
where $\mathcal{A}_U$ denotes the von Neumann algebra associated with the subregion $U$. In the cases of local modular flows, this property is manifested geometrically by that the modular trajectories through any point $x\in D[U]$ remain inside $D[U]$, where $D[U]$ is the casual wedge of of $U$. 

\begin{figure}[h]
    \centering
    \subfloat[$\gamma=0$]{\includegraphics[width=0.5\linewidth]{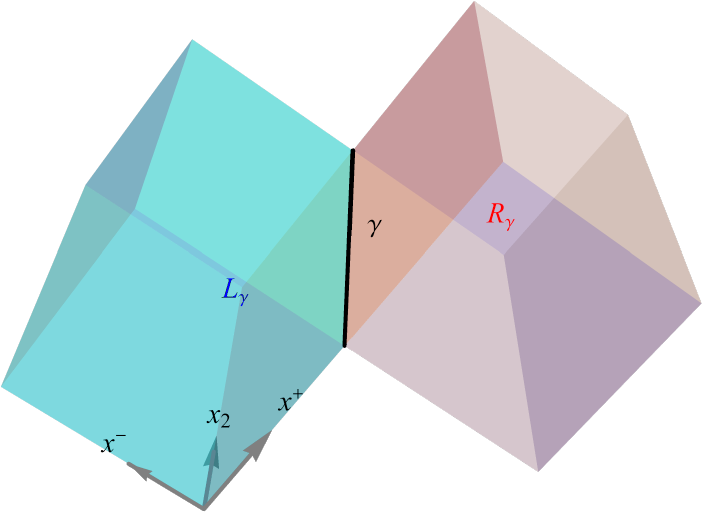}}\hspace{2ex}
    \subfloat[$\gamma=(x_2)^2$]{\includegraphics[width=0.4\linewidth]{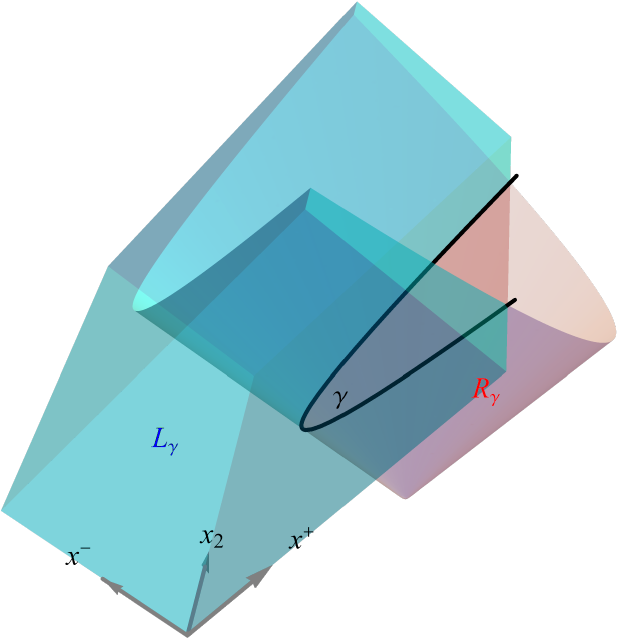}}
    \caption{Right: the causal wedges $\mathcal{R}_{\gamma}$ and $\mathcal{L}_{\gamma}$ across $\gamma(x_2)=(x_2)^2$; Left: comparison with the Rindler wedges ($\gamma=0$) to help with visual perception.}
    \label{fig: z2square}
\end{figure}

Being a general statement, the Tomita-Takesaki theory should remain valid in the non-local case. It is therefore worth checking its prediction against our results. Doing this can both serve as a consistency check and possibly reveal novel features. In what follows, we focus on the explicit non-local modular flow from the $\gamma(x_2)=(x_2)^2$ as an example. More specifically, we study the modular commutator:
\be\label{eq:squared_nonlocal}
[\hat{H}_\gamma,\phi(x)] = -i\pi m(x^-)^2 \partial_{x^-} \int \dd z_2\ e^{-m|x_2-z_2|}\phi(x^+,x^-,z_2) +\cdots,
\ee
where we omit the local terms. This is an integral of local operators supported on the unbounded one-dimensional space-like sub-manifold $S_x = \lbrace (x^+,x^-,z_2)\,|\,z_2\in\mathbb{R} \rbrace$. One can compute the right and left causal wedges across $\gamma$. They are defined by the following inequalities, see Figure \ref{fig: z2square}: 
\begin{align}
        \mathcal{R}_{\gamma}&=\{x\in\mathbb{R}^{1,2}\,|\,-x^+(x^-+1)+(x_2)^2<0,\ -1<x^-<0\},\\
        \mathcal{L}_{\gamma}&=\{x\in\mathbb{R}^{1,2}\,|\,-x^+(x^-+1)+(x_2)^2>0,\ x^->0\}.
\end{align}
In particular, for $x = \left(x^+,x^-,x_2\right)\in \mathcal{R}_\gamma$, one can check that the non-local support $S_x$ of the modular commutator, though never enters the complement causal wedge $\mathcal{L}_\gamma$, always contains a part that lies outside $\mathcal{R}_\gamma$:
\be
S_x \setminus {\mathcal{R}_\gamma} = \lbrace (x^+,x^-,z_2)\,|\, (z_2)^2 > x^+(x^-+1)\rbrace.
\ee

It may look like that the part of $[\hat{H}_\gamma,\phi(x)]$ supported in $S_x \setminus {\mathcal{R}_\gamma}$ is in the causal influence of both $\mathcal{R}_\gamma$ and $\mathcal{L}_\gamma$. As a result $[\hat{H}_\gamma,\phi(x)]$ may have non-zero overlap with the von Neumann algebra of $\mathcal{L}_\gamma$, i.e. the infinitesimal modular flow may move $\phi(x)$ beyond $\mathcal{R}_\gamma$. We emphasize that a closer examination reveals that this does not happen, and the Tomita-Takesaki theory remains consistent with our results. To clarify this subtlety, in the \textbf{Appendix} \ref{app:modular_flow_causality} we perform a detailed computation of the commutator between the infinitesimally modular flowed operator $[\hat{H}_\gamma,\phi(x)]$ for any $x\in \mathcal{R}_\gamma$  and the local operator $\phi(y)$ for any $y\in \mathcal{L}_\gamma$. We show that despite the possibility of non-trivial causal influence between the support of the two operators, their commutator vanishes identically in a subtle way. In other words, $[\hat{H}_\gamma,\phi(x)]$ remains an element of the von-Neumann algebra of $\mathcal{R}_\gamma$, consistent with the Tomita-Takesaki theory. 

\section{Finite modular flows}\label{sec:flow}

Having obtained the modular commutator $[\hat{H}_{\gamma},\phi(x)]$,
this section proceeds to investigate the finite modular flow generated by $\hat{H}_{\gamma}$.
Our goal is to carry out the computation as explicitly as possible. Doing this for the general null-cut $\gamma(x_2)$ involves too much technicalities and is beyond the scope of this work. For this reason, we choose to focus on the specific case of $\gamma(x_2) =\alpha (x_2)^2$ with $m\neq 0$. In particular, we expect that the coefficient $\alpha$ can mimick the local curvature of a general cut $\gamma$ on the null-plane -- especially when the operator is close to it.  As was observed, this is the simplest case where the non-local aspect of modular flow begins to show up. Admittedly, considering modular flows across such a particular null-cut boundary in a special (free) theory is far from being general. The hope is that by studying it we can learn important aspects of non-local modular flows that are universally shared by the more general cases, e.g. when and how can we approximate the modular flow by the corresponding local boost for operators close to the entangling boundary.

We make an additional remark that this set-up is in spirit similar to the case of non-local modular flow associated with the spherical subregion in the vacuum of a massive theory, which also becomes local in the CFT (massless) limit. A crucial distinction between the two cases is that our set-up features the additional property of half-sided modular inclusion, so the explicit form of the modular Hamiltonian is known for all masses. This is not available for the case of spherical subregion in a massive vacuum, whose non-local modular flow is therefore more difficult to study.   

\begin{figure}[h]
    \begin{center}
    \includegraphics[scale=0.9]{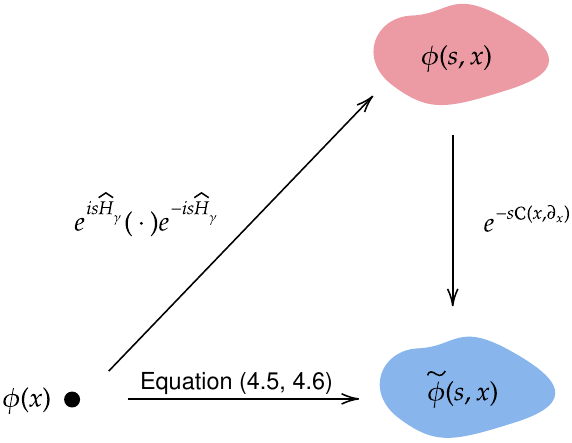}
    \end{center}
    \caption{The modular flowed operator $\phi(s,x)$ is obtained by the adjoint action of the evolution operator $\exp(-is \hat{H}_{\gamma})$ on the initial value $\phi(x)$. Then we can strip off the local flow from the $\phi(s,x)$ by acting with the inverse local evolution operator $\exp\left(-s\mathcal{C}(x,\partial_x)\right)$ on it. The defining equations for $\tilde{\phi}(s,x)$ are given by \eqref{eq:tilde phi evolution} \eqref{eq:tilde phi initial}.}
    \label{fig: commutative diagram}
\end{figure}

To obtain the modular flow equation of $\phi(s,x)$,
we act modular evolution operator $e^{-is\hat{H}_{\gamma}}$ adjointly on both sides of the modular commutator $[\hat{H}_{\gamma},\phi(x)]$.
For the case of $\gamma(x_2) = \alpha (x_2)^2$,
$P_{\alpha (x_2)^2}=\alpha P_{ (x_2)^2}$,
equation \eqref{eq:gamma=z2square} gives us
\begin{equation}\label{eq: phi evolution}
        \partial_s\phi(s,x)=\mathcal{C}(x,\partial_x)\phi(s,x)+\alpha \pi m(x^-)^2\partial_{x^-}\int\dd y_2\ e^{-m|x_2-y_2|}\phi(s,x^0,x_1,y_2),
\end{equation}
where we use $\mathcal{C}(x,\partial_x)$ to denote the local part,
\begin{equation}
        \mathcal{C}(x,\partial_x)=2\pi \left[\left(x^+-\alpha(x_2)^2\right)\partial_{x^+}-x^-(1+\alpha x^-)\partial_{x^-}-\alpha x^-x_2\partial_{x_2}-\frac{\alpha}{2}x^-\right].
\end{equation}
If the mass is equal to zero, the equation \eqref{eq: phi evolution} becomes a first-order partial differential equation, which can be systematically solved by transforming it into the first integrals of ordinary differential equations. Specifically, by ignoring the coefficient function, we obtain  $\phi(s,x)\propto\phi(x(s))$ where the trajectory $x(s)$ is:
\begin{eqnarray}
	x^+\left( s \right) &=&\frac{e^{4\pi s}\left( x^+-\alpha \left( x_2 \right) ^2+\alpha x^+x^- \right) +\alpha e^{2\pi s}\left( \left( x_2 \right) ^2-x^+x^- \right)}{e^{2\pi s}\left( \alpha x^-+1 \right) -\alpha x^-}\nonumber,\\
	x^-\left( s \right) &=&\frac{x^-}{e^{2\pi s}\left( \alpha x^-+1 \right) -\alpha x^-}\nonumber,\\
	x_2\left( s \right) &=&\frac{e^{2\pi s}x_2}{e^{2\pi s}\left( \alpha x^-+1 \right) -\alpha x^-}.
\end{eqnarray}
Therefore, we focus our attention on the effects of the non-local term in \eqref{eq: phi evolution}. Analogous to the interaction picture approach, we can remove the local evolution of $\phi(s,x)$ by introducing the transformed field, see figure \ref{fig: commutative diagram}:
\begin{equation}
    \tilde{\phi}(s,x)=e^{-s\mathcal{C}(x,\partial_x)}\phi(s,x).
\end{equation}
By stripping-off the effect of the local action, we can derive the following evolution equation for $\tilde{\phi}(s,x)$ that only contains the non-local effects:
\begin{equation}\label{eq:tilde phi evolution}
    \begin{aligned}
        \partial_s\tilde{\phi}(s,x)&=\alpha \pi me^{2\pi s}(x^-)^2\left[\partial_{x^-}+\alpha\frac{e^{2\pi s}-1}{(e^{2\pi s}-1)\alpha x^--1}(x_2\partial_{x_2}+\frac{1}{2})\right]\\
        &\times \int\dd y_2\ \exp\left(-m|e^{2\pi\alpha sx^-}x_2-y_2|\right)\tilde{\phi}(s,x^+,x^-,y_2).
    \end{aligned}
\end{equation}
This is a differential-integral equation. To describe the modular flow of initial local operators, we also supply (\ref{eq:tilde phi evolution}) with the initial condition for $\tilde{\phi}(s,x)$:
\begin{equation}\label{eq:tilde phi initial}
    \tilde{\phi}(0,x)=\phi(x).
\end{equation}

In a free theory of scalar $\phi$, the local  operator $\phi(x)$ and the conjugate momentum $\pi(x)$ for all $x$ on a Cauchy surface form a complete operator basis. A useful property of the evolution equation (\ref{eq:tilde phi evolution}) is that the right-hand-side does not contain derivative or integral along the $x^+$ direction. For an initial local operator $\phi(x)$, the evolved operator $\tilde{\phi}(s,x)$ remains fixed in the $x^+$ direction -- the fully-flowed operator $\phi(s,x)$ does move and spread along $x^+$, but that only comes from the action of local evolution operator $\exp\left(s\mathcal{C}(x,\partial_x)\right)$. In other words, $\tilde{\phi}(s,x)$ is supported on the null surface $\lbrace y\in \mathbb{R}^{1,2}\,|\,y^+=x^+\rbrace $, and can therefore be expanded by local operators on the null surface, which allows us to rewrite it in terms of a kernel function $G$:
\begin{equation}\label{eq: ansatz}
    \tilde{\phi}(s,x)=\int_{(x^+,y^-,y_2)\in \mathcal{R}_{\gamma}}\dd y^-\dd y_2\ G(s,x^-,x_2,y^-,y_2)\phi(x^+,y^-,y_2),
\end{equation}
where we remind that the conjugate momentum contribution has been include by the derivative $\partial_-$.  Substituting the ansatz \eqref{eq: ansatz} into the equation \eqref{eq:tilde phi evolution},
we derive the following differential-integral equation for the kernel function $G$:
\begin{align}\label{eq:G evolution}
    \partial_sG(s,x^-,x_2,z^-,z_2)&=\alpha\pi me^{2\pi s}(x^-)^2\left[\partial_{x^-}+\alpha\frac{e^{2\pi s}-1}{(e^{2\pi s}-1)\alpha x^--1}(x_2\partial_{x_2}+\frac{1}{2})\right]\notag\\
    &\times\int\dd y_2\  \exp\left(-m|e^{2\pi\alpha sx^-}x_2-y_2|\right)G(s,x^-,y_2,z^-,z_2),
\end{align}
supplied with the initial condition corresponding to 
\eqref{eq:tilde phi initial}:
\begin{equation}\label{eq:G initial}
    G(0,x^-,x_2,z^-,z_2)=\delta(x^--z^-)\delta(x_2-z_2).
\end{equation}
\subsection{Perturbation theory in small $x^-$}
We should not expect to solve the differential-integral equation (\ref{eq:G evolution}) analytically. On the other hand, our goal is to study non-local modular flows that remain close to the entangling surface, in this case characterized by small $x^-$. Most of the intuitions regarding generic modular flows applies only in this limit. Based on the observation that the non-local terms in the modular flow, hence the RHS of (\ref{eq:G evolution}), is suppressed by small $x^-$, we expect that a perturbation theory can be organized regarding the solution to (\ref{eq:G evolution}). To proceed, 
we expand the kernel function $G$ in powers of $x^-$:
\begin{equation}\label{eq: G expansion}
    G(s,x^-,x_2,z^-,z_2)=\sum_{n=0}^{\infty}(x^-)^n G_{(n)}(s,x_2,z^-,z_2)
\end{equation}
The proximity to the entangling surface is reflected in the choice of the initial condition (\ref{eq:G initial}), which can also be expanded in powers of small $x^-$ and becomes the corresponding initial conditions for the $G_{(n)}$ order by order:
\begin{align}
    G_{(0)}(0,x_2,z^-,z_2)&=\delta(z^-)\delta(x_2-z_2),\\
    G_{(1)}(0,x_2,z^-,z_2)&=-\partial_{z^-}\delta(z^-)\delta(x_2-z_2),\label{eq: G1 initial condition}\\
    G_{(2)}(0,x_2,z^-,z_2)&=\frac{1}{2}\partial^2_{z^-}\delta(z^-)\delta(x_2-z_2). \label{eq: G2 initial condition}
\end{align}
In this framework, the evolution equation \eqref{eq:G evolution} can be expanded order by order and solved iteratively. The first few orders can be written explicitly as:
\begin{align}
    \partial_sG_{(0)}(s,x_2,z^-,z_2)&=0,\\
    \partial_sG_{(1)}(s,x_2,z^-,z_2)&=0,\\
    \partial_sG_{(2)}(s,x_2,z^-,z_2)&=\alpha\pi me^{2\pi s}\left[(-2\pi s+e^{2\pi s}-1)\alpha mx_2\mathrm{sgn}(x_2-z_2)-\partial_{z^-}-\alpha\frac{e^{2\pi s}-1}{2}\right]\notag\\
    &\times e^{-m|x_2-z_2|}\delta(z^-).
\end{align}
Supplied with the corresponding initial conditions (\ref{eq:G initial}), the differential equations in $s$ can be solved explicitly. The solutions for the first few orders are given by:
\begin{align}
    G_{(0)}(s,x_2,z^-,z_2)&=\delta(z^-)\delta(x_2-z_2),\\
    G_{(1)}(s,x_2,z^-,z_2)&=-\partial_{z^-}\delta(z^-)\delta(x_2-z_2),\\
    G_{(2)}(s,x_2,z^-,z_2)&=\Big[\frac{m^2\alpha^2}{4}x_2(e^{4\pi s}-4\pi se^{2\pi s}-1)\mathrm{sgn}(x_2-z_2)\notag\\
    &-\frac{m\alpha}{2}(e^{2\pi s}-1)\partial_{z^-}-\frac{m\alpha^2}{8}(e^{2\pi s}-1)^2\Big]e^{-m|x_2-z_2|}\delta(z^-).
\end{align}
We can recursively derive higher order corrections in the same way. This process will become inevitably more complicated,
so we truncate it at the second order. Notice that the perturbative solution is written in terms of the coefficient functions that are proportional to $\delta(z^-)$, i.e. supported only on the $z^-=0$ surface.  

From the perturbative solution of the kernel function $G$, we can write down the flowed operator $\tilde{\phi}(s,x)$ expanded in small $x^-$ up to second order:
\begin{eqnarray}\label{eq: tilde phi 2 orders}
            \tilde{\phi}(s,x)&=&\left(1+x^-\partial_-+\frac{(x^-)^2}{2}\partial_-^2\right)\phi(x^+,0,x_2)\nonumber\\
            &+&\frac{m^2\alpha^2}{4}(x^-)^2x_2(e^{4\pi s}-4\pi se^{2\pi s}-1)\int\dd z_2\ e^{-m|x_2-z_2|}\mathrm{sgn}(x_2-z_2)\phi(x^+,0,z_2)\nonumber\\
            &+&\frac{m\alpha}{8}(x^-)^2(e^{2\pi s}-1)\left(4\partial_{x^-}|_{x^-=0}-\alpha e^{2\pi s}+\alpha\right)\int\dd z_2\ e^{-m|x_2-z_2|}\phi(x^+,0,z_2)\nonumber\\
            &+&...   \end{eqnarray}
The term in the first line is merely Taylor expansion of $\phi(x)$ at $x^-=0$,
while the term in the second and third line represent the non-local behavior of the modular flow,
which is a convolution with respect to the kernel function in $x_2$.

\subsection{Re-summation for $s \to \infty$ at fixed $\lambda =\alpha (x^-) e^{2\pi s}$}
At general orders of $(x^-)^n$, $G_{(n)}(s,x_2,z^-,z_2)$ contains terms that grow exponentially with $s$.
It can be observed that the fastest growing factor term is proportional to $(\alpha x^-)^ne^{2\pi n s}$. For this reason, a natural large $s$ limit emerges as taking $\left(x^-\to 0,\;s\to \infty\right)$, while keeping the combination $\lambda=\alpha (x^-)e^{2\pi s}$ fixed. In this limit  $G(s,x^-,x_2,z^-,z_2)$ is dominated by the terms proportional to $e^{2\pi ns}$ in $G_{(n)}(s,x_2,z^-,z_2)$ at each order $n$.
Let us define the factor $g_n$ for the fastest growing terms in $G_{(n)}$: 
\be
G_{(n)}(s,x_2,z^-,z_2) = \lambda^{n}g_{(n)}(x_2,z^-,z_2) + \cdots,
\ee
where $\cdots$ denotes the remaining terms that grow slower than $e^{2\pi n s}$. The full kernel function in this limit is then given by: 
\be
G(s,x^-,x_2,z^-,z_2) \approx \sum_n \lambda^n g_{(n)}(x_2,z^-,z_2),\;\;\lambda = \alpha\left(x^-\right) e^{2\pi s}.
\ee
We can explore the non-perturbative features of $G$ that emerge after re-summing the $\lambda$ series. To begin with, we can derive the following recursion equations:
\begin{equation}\label{eq:g_recursive}
    \begin{aligned}
        &2(n+1) g_{(n+1)}(x_2,z^-,z_2)=n m \int\dd y_2\ e^{-m|x_2-y_2|} g_{(n)}(y_2,z^-,z_2)\\
        &- m \left(x_2\partial_{x_2}+\frac{1}{2}\right)\int\dd y_2\;e^{-m|x_2-y_2|}\sum_{k=0}^{n-1} g_{(k)}\left(y_2,z^-,z_2\right).
    \end{aligned}
\end{equation}
Solving (\ref{eq:g_recursive}) in exact terms is still too difficult. On the other hand, the main features of the re-summed effects can usually be captured by contributions at large orders $n$, for which the RHS of (\ref{eq:g_recursive}) is dominated by the first term \footnote{Strictly speaking, the remaining part on the RHS of (\ref{eq:g_recursive}) involves a sum over $n$-terms, which could collectively contribute at the same order as the first term at large $n$. We acknowledge this and proceed under the assumption that considering only the first term should at least qualitatively capture the re-summed effects. We leave a more careful treatment for future investigations. 
For now we can perform a crude self-consistency check by substituting the resulting solution $g_{(n)}$ \eqref{eq:resum} back into \eqref{eq:g_recursive} and estimating the large $n$-scaling of the neglected terms:
\begin{equation}
\begin{aligned}
 \sum_{k=0}^{n-1}g_{(k)}(x_2,z^-,z_2)&=e^{-m|x_2-z_2|}\sum_{k=0}^{n-1}\left[\frac{1}{4^k}\left( \begin{array}{c}
	2k\\
	k\\
\end{array} \right) +\cdots+\left(\frac{m}{2}\right)^k\frac{|x_2-z_2|^k}{k!}\right].
\end{aligned}
\end{equation}
Similar to the main analysis, we can sandwich the full expression by only considering the two limiting terms, whose large $n$-scalings are given by:
\begin{eqnarray}
\sum_{k=0}^{n-1} \frac{1}{4^k} \left(\begin{array}{c}
	2k\\
	k\\
\end{array}\right) \lesssim \sum^{n-1}_k \frac{1}{\sqrt{\pi k}} \sim O(\sqrt{n}),\;\;\;\sum^{n-1}_{k=0} \left(\frac{m}{2}\right)^k\frac{|x_2-z_2|^k}{k!} \lesssim O(1)  .
\end{eqnarray}
We see that both terms remain sub-dominant to the leading-order $O(n)$ term in (\ref{eq:g_recursive}). 
}: 
\be
 g_{(n+1)}(x_2,z^-,z_2) =  \frac{ m}{2} \int dy_2 e^{-m|x_2-y_2|} g_{(n)}(y_2,z^-,z_2) + \mathcal{O}(n^{-1}) .
\ee
The solution at the leading order in large $n$ involves an $n$-dimensional convolution integral, which can be evaluated to yield the general form:
\be
\int \prod^{n}_{i=1}dw_i\; e^{-m|x-w_{n}|}e^{-m|w_n-w_{n-1}|}...e^{-m|w_1-y|} = m^{-n}e^{-m|x-y|} P_n\left(m|x-y|\right).
\ee
The factor $P_n(x)$ is a degree-$n$ polynomial. For illustration we list explicitly $P_n(x)$ for the first few orders:
\be 
P_1(x) = 1+x,\;\;P_2(x)=\frac{3}{2}+\frac{3x}{2}+\frac{x^2}{2},\;\;P_3(x) = \frac{5}{2}+\frac{5}{2}x+x^2+\frac{x^3}{6}.
\ee
At general $n$ the polynomial $P_n(x)$ has simple expressions for the $x^0$ and $x^n$ terms: 
\be
P_n(x) =\frac{(2n)!}{2^n(n!)^2}+\cdots+\frac{x^n}{n!}.
\ee
This then allows us to perform the following (partial) re-summation:
\bea\label{eq:resum}
G(s,x^-,x_2,z^-,z_2) &\propto & e^{-m|x_2-z_2|} \sum_n \left\lbrace \left(\frac{ \lambda}{4}\right)^n \frac{(2n)!}{(n!)^2}+...+\frac{( m\lambda)^n |x_2-z_2|^n}{2^n n!} \right\rbrace \nonumber\\
&\sim & \frac{e^{-m|x_2-z_2|}}{\sqrt{1-\lambda}}+...+e^{m(\frac{\lambda}{2}-1)|x_2-z_2|},
\eea
where we have omitted writing the $z_1$ dependent factor. The two explicit terms represent the limiting behaviours of the re-summed kernel function $G$: the first term dominates for nearby points, i.e. those with $m|x_2-z_2|\ll 1$; the last term dominates for asymptotically separated points, i.e. those with $m|x_2-z_2|\gg 1$. The $\cdots$ represent intermediate terms, and we may think of the full result of (\ref{eq:resum}) as being ``sandwiched" by the two limits. For this reason our discussion of (\ref{eq:resum}) will focus only on its first and last terms.
  
Let us re-emphasize that (\ref{eq:resum}) becomes valid only after the following two limits have been taken: (i) leading order in $x^-\to 0, s\to \infty$ at fixed $\lambda =\alpha (x^-)e^{2\pi s}$; (ii) leading order in $n^{-1}$ for the large order contributions. The latter limit only extracts the leading order singularity or phase transition of $G$ as a function of $\lambda$. It is only in this sense that we should treat (\ref{eq:resum}) as a valid approximation to the exact kernel function. Such singularities or phase transitions are in fact manifested in (\ref{eq:resum}). The first term -- representing $G$ at the nearby points -- encounters a singularity as $\lambda$ increases across $1$; the last term -- representing $G$ at large separations -- changes from exponential decaying to exponential growing as $\lambda$ increases across $2$. Both observations imply the breakdown of the perturbation theory, and consequently the invalidity of approximating the modular flow by local trajectories,
at $\lambda \sim \mathcal{O}(1)$.
For fixed $0<\alpha x^- \ll 1$, this can be interpreted as providing an upper limit $\bar{s}$ for the modular time $s$:
\be
\bar{s}\sim -\frac{1}{2\pi}\ln{\left(\alpha x^{-}\right)}.
\ee
Only for $s\lesssim \bar{s}$ can the modular flow be approximated by trajectories that are ``adiabatically" smeared out; while for larger $s$ this picture breaks down. Roughly speaking, the breakdown described by the first term in (\ref{eq:resum}) corresponds to the development of ``caustics" among nearby points \footnote{This should not be identified with the potential caustics formed among the local trajectories described by the local terms in the modular generators -- those motions have been stripped off in $\tilde{\phi}(s,x)$. For $\gamma(x)=x^2$ the local trajectories are integral curves generated by conformal isometries and there are no caustics except at the tips of the causal domain.}; the breakdown described by the last term in (\ref{eq:resum}) corresponds to the modular flow becoming fully spread-out -- the kernel function $G$ is not suppressed at large separations.

We end this section by making the observation that depending on the sign of the curvature $\alpha$, the actual singularity for $\lambda$ occurs only on a particular side of the entangling surface. For $\alpha>0$ it occurs on the side $x^->0$; for $\alpha<0$ it occurs on the side $x^-<0$. Such qualitative distinction between the two sides under modular flow is a novel feature for the curved entangling surface. The asymmetry can also be revealed by comparing the causal developments for the two sides, see figure \ref{fig: z2square}.

\section{Discussions}\label{sec:discussion}
In this paper, we study the modular flows associated with relativistic vacuum state across entanglement boundaries $x^{\bot}=\gamma(x^{\bot})$ that lie on the null plane $x^-=0$. Due to the algebraic property of the half-sided modular inclusion in these cases, the full modular Hamiltonians $\hat{H}_\gamma$ can be written explicitly in terms of integrals of the stress tensor operator $T_{++}$ weighted by $\gamma$. We are primarily interested in studying the non-local aspects of the modular flows, i.e. the way in which the modular flow of local operator $\mathcal{O}(s,x) = e^{iH_\gamma s}\mathcal{O}(x)e^{-iH_\gamma s}$ cannot be identified with an operator localized in space-time. 

By focusing on the free QFT of massless or massive scalar $\phi$ in 1+2 dimension, the modular flow generator $[\hat{H}_\gamma,\phi(x)]$ can be explicitly computed in exact terms. The results can be organized by decomposing the profile function into  monomials $\gamma(x_2) = (x_2)^n$. For the cases of linear profile functions $\gamma(x_2) \in\lbrace 1, x_2\rbrace$ the modular flows are local as a result of the relativistic symmetry; for the quadratic case  $\gamma(x_2)= (x_2)^2$ the modular flow is still local for the massless field due to the underlying conformal symmetry, and becomes non-local in the massive case; for higher powers $\gamma(x_2) = (x_2)^n,n\geq 3$ the modular flows are non-local in both the massive and massless cases. General patterns of local and non-local terms in the modular commutators for arbitrary powers $n$ are identified and summarized in \eqref{eq: general non-local}.  

We therefore choose to focus on the simplest non-local case in this class: a massive free scalar across quadratic null-cuts: $\gamma(x_2) = \alpha  (x_2)^2$. Using explicit results for the modular commutator, we derive integral-differential equations \eqref{eq:tilde phi evolution} and \eqref{eq:tilde phi initial} satisfied by
``non-local'' part of the corresponding modular flowed operator $\tilde{\phi}(s,x)$. The equations can be solved in perturbation theory of the distance $x^-$ deviating from the null plane. The leading two orders are computed in \eqref{eq: tilde phi 2 orders}. Motivated by the general form of the expansion, we focused further on the limit of $x^-\to 0, s\to \infty$ keeping $\lambda = \alpha(x^-)e^{2\pi s}$ fixed, in which the modular flowed operator (represented in the free theory by the kernel function $G$) can be expanded in power series of the combination $\lambda$. For the purpose of extracting the leading order singularities or phase transitions in $\lambda$, the series can be (partially) re-summed into the explicit form (\ref{eq:resum}). The result points to a breakdown of perturbation theory, and hence the invalidity of the adiabatic approximation of $\tilde{\phi}(s,x)$ by smeared local operators, at $\lambda \sim \mathcal{O}(1)$. The breakdown can either be diagnosed by the formation of caustics among nearby points, or by the non-decaying behavior of the kernel function at large separations.   

Despite the non-local nature of modular flows in the general cases, in practice what underlies the discussions in most contexts is the intuition that generic modular flows can be approximated by local Rindler boosts very close to the entangling surface. However, explicit details regarding this intuition remain obscure. Further questions that can be asked include: what controls the approximation behind this intuition; what is the regime of validity for the intuition; and how does it breakdown? Qualitative speculations can be proposed based on general principles. Our results provide an opportunity to examine these questions in concrete terms.  In what follows we enumerate and discuss the relevant features reflected by our results.
\begin{itemize}[leftmargin=*]
\item The instantaneous modular generator on local operators $[\hat{H}_\gamma, \phi(x)]$ contains a local term and a non-local term. \item The local terms are differential operators acting on $\phi(x)$. They can be understood as specifying the vector fields whose integral curves correspond to the would-be trajectories that the approximation is based on.  In the general scenarios without (conformal) isometries, these trajectories will encounter caustics.   
\item The non-local terms are characterized by an integral-convolution in the longitudinal $x_2$ direction, acted further by differential operators in the transverse $x^-$ direction. This reveals the instantaneous non-local flow as the simultaneous motions of both spreading along the $x_2$ direction and transporting along the $x^-$ direction. 
\item The non-local terms are all proportional to the instantaneous transverse distance $x^-$ to the null plane, and on the null plane $x^-=0$, the non-local terms remain vanishing and the full modular flows reduce to null translations. These are fully in support of the aforementioned intuition regarding generic modular flows near the entangling boundary.  
\item For $x$ in one of the causal wedges across $\gamma$, e.g. $x\in \mathcal{R}_\gamma$, it is possible for the non-local term in $[\hat{H}_\gamma,\phi(x)]$ to be supported outside $\mathcal{R}_\gamma$. However, $[\hat{H}_\gamma,\phi(x)]$ still commutes with all operators in the von-Neumann algebra of $\mathcal{L}_\gamma$, and therefore remains an element of the von-Neumann algebra of $\mathcal{R}_\gamma$.   
\item Away from the null plane, the modular flowed operator satisfies an integral-differential equation after stripping away the effect of the local terms in the modular commutator. After solving it in perturbation theory of the small initial transverse distance $x^-$, the solution admits a natural series expansion in $\lambda = \alpha x^- e^{2\pi s}$. The parameter $\lambda$ is in fact what controls the approximation underlying the intuition, which suffers a breaks down at $\lambda \sim \mathcal{O}(1)$. 
\item In terms of the modular time $s$, the perturbation theory is valid for $s\lesssim -\ln{\left(\alpha x^-\right)}$. We can thus conclude that the closer to the entangling boundary the initial operator is located, or the smaller the curvature of the entangling surface is, the longer time $s$ under modular flow the Rindler approximation remains valid.    
\item The breakdown is primarily due to the spreading along the longitudinal $x_2$ direction. Its effects towards $\lambda \to \mathcal{O}(1)$ can be analyzed in two limits. Far from the center of support, it undermines the approximation of $\tilde{\phi}(s,x)$ as a local operator by making the weight kernel $G$ non-decaying; near the center, it creates caustics in the form of singularities for the weight kernel $G$.

\item It is worth noting that the mass coupling $m$, though crucial for the modular flow to be non-local for $\gamma(x_2)= (x_2)^2$ and therefore controls the magnitude of non-local effects, nevertheless does not enter the parameter $\lambda$ that controls the validity of perturbation theory.   
\end{itemize} 

We conclude the paper by discussing a few future directions. Though we hope our results capture the general features of non-local modular flows, the contexts in which they are derived are not fully general. Further modifications can be considered such as excitation above the vacuum, interacting QFTs, or more general entanglement boundaries, etc. In the future it is important to extend such analysis to contexts of increasing generalities. At a more technical level, the re-summation in the limit of $x^-\to 0$ at fixed $\lambda$ could be treated in a manner that is more careful and thorough. It is worth examining in future works whether doing so will reveal more interesting results. Regarding the near-null-plane perturbation theory in small $\alpha$, many of its aspects require further refinements, for example: how to quantitatively construct an effective ``distance measure" for the resemblance of the general modular flow with respect to a local boost when $\alpha\ll 1$; the current form of perturbation theory only applies near the future null boundary $x^-=0$, modular flows near the past null boundary are drastically distinct from the future counter-part, how to describe them is another direction for the future.
On the other hand, in the face of the breakdown of perturbation theory at $\lambda \sim \mathcal{O}(1)$, a natural question for future works is what characterizes the modular flow for larger values of $\lambda$, for example for $\lambda \gg 1$. At fixed $x^-$, this is probing the long time response of modular dynamics near modular equilibrium, and it will be very interesting if novel features absent in conventional thermodynamics can be discovered. Besides, based on our results regarding non-local modular flows in the vacuum state $\Omega$ across deformed null cuts $\gamma$, an interesting endeavor in the future is to explicitly reverse-engineer a corresponding deformation of the state $\Omega \to \Psi_\gamma$ such that the non-local effects are cancelled, giving rise to local modular flow. Such constructions have played important roles in the recent progresses about Type II von-Neumann algebra from cross product, which characterizes perturbative quantum gravity \cite{Witten_2022,Chandrasekaran_20231,Chandrasekaran_20232,Kudler-Flam:2023qfl,Jensen:2023yxy}. 

\section*{Acknowledgment}
Guan-Cheng Lu would like to thank Wen-Xin Lai for helpful discussions. This work is supported by National Science Foundation of China (NSFC) grant no. 12175238.

\section{Appendix}

\subsection{Computation of $[P_{\gamma},\phi(x)]$ for $\gamma(z_2)=(z_2)^2$}\label{sec:appendix z2square}
In this appendix, we will compute the commutator $[P_{\gamma},\phi(x)]$ for the null-cut $\gamma(z_2)=(z_2)^2$.
The calculation process is same for any null-cut $\gamma(z_2)=(z_2)^n$,
although the details may be complicated.
Substitute $\gamma(z_2)=(z_2)^2$ into the equation \eqref{eq:A},
we have
\begin{equation}
    \begin{aligned}
       A|_{y^0=x^0}=\frac{1}{8}\int\frac{\dd ^2p}{(2\pi)^2}&e^{i\vec{p}\cdot(\vec{x}-\vec{y})}\Big\{\left[(x^0-x_1)^2+(x^0-y_1)^2+(x_2)^2+(y_2)^2\right]\\
       &-m^2\frac{(x^0-x_1)^2+(x^0-y_1)^2}{m^2+(p_2)^2}\Big\}+i(\text{real function}).
    \end{aligned}
\end{equation}
Here, $ i(\text{Real function})$ is the summation of terms as follows,
\begin{equation}\label{eq: imaginary part}
    \int\dd^2p\ e^{i\vec{p}\cdot(\vec{x}-\vec{y})}\frac{i^{\alpha}(p_1)^{\beta}(p_2)^{\gamma}}{(m^2+(p_2)^2)\sqrt{(p_1)^2+(p_2)^2+m^2}},
\end{equation}
where $\alpha,\beta,\gamma$ are non-negative integers and $\alpha+\beta+\gamma$ is odd.
To find that the integral is pure imaginary,
we replace $p_j e^{i\vec{p}\cdot(\vec{x}-\vec{y})}$ with $\partial/(i\partial x_j)e^{i\vec{p}\cdot(\vec{x}-\vec{y})}$,
\begin{align}
    \mathrm{Re}[\eqref{eq: imaginary part}]&\propto\frac{\partial^{\beta}}{(\partial x_1)^{\beta}}\frac{\partial^{\gamma}}{(\partial x_2)^{\gamma}}\mathrm{Im}[\int\dd^2p\ \frac{e^{i\vec{p}\cdot(\vec{x}-\vec{y})}}{(m^2+(p_2)^2)\sqrt{(p_1)^2+(p_2)^2+m^2}}]\notag\\
    &\propto \frac{\partial^{\beta}}{(\partial x_1)^{\beta}}\frac{\partial^{\gamma}}{(\partial x_2)^{\gamma}}\int_0^{+\infty}\frac{\dd p}{(2\pi)^2}\int_0^{2\pi}\dd\theta\ 
    \frac{p\sin\left[ip(x_1-y_1)\cos\theta+ip(x_2-y_2)\sin\theta\right]}{\left(m^2+p^2\sin^2\theta\right)\sqrt{m^2+p^2}}\notag\\
    &=0.
\end{align}
In the last step, we have utilized the following integral,
\begin{equation}
    \int_0^{2\pi}\dd\theta\ \frac{\sin\left[a\cos\theta+b\sin\theta\right]}{(c+\sin^2\theta)^n}=0,\quad n\in\mathbb{Z}_{\ge 0}.
\end{equation}
After integrating out $p$,
we obtain,
\begin{equation}\label{eq:A+A* gamma=z2square}
    A+A^*|_{y^0=x^0}=\frac{1}{2}\left[(x^-)^2+(x_2)^2\right]\delta^2(\vec{x}-\vec{y})-\frac{1}{4}m(x^-)^2\delta(x_1-y_1)e^{-m|x_2-y_2|}.
\end{equation}
Similarly, by taking $\gamma(z_2)=(z_2)^2$ in equation \eqref{eq:B},
we have
\begin{align}
        & B|_{y^0=x^0}=\frac{1}{8}\int\frac{\dd^2 p}{(2\pi)^2}e^{i\vec{p}\cdot(\vec{x}-\vec{y})}\Big[
            ip_1\left((x^-)^2+(x^0-y_1)^2-(x_2)^2-(y_2)^2\right)+2x^0+x_1-3y_1\notag\\
            &+2ip_2\left(-x^0(x_2+y_2)+x_1x_2+y_1y_2\right)-m^2\frac{ip_1\left((x^-)^2+(x^0-y_1)^2\right)+2x^0-2y_1}{m^2+(p_2)^2}\Big]\notag\\
        &+i(\text{real function}).
\end{align}
Here, we have employed the same computational techniques as above,
which yields,
\begin{equation}\label{eq:B+B* gamma=z2square}
    \begin{aligned}
        B+B^*|_{y^0=x^0}&=\frac{1}{2}\left[\left((x^-)^2-(x_2)^2\right)\partial_{1}-2x^-x_2\partial_{2}-x^-\right]\delta^2(\vec{x}-\vec{y})\\
        &-\frac{1}{4}m(x^-)^2\partial_{x_1}\delta(x_1-y_1)e^{-m|x_2-y_2|}.
    \end{aligned}
\end{equation}
Ultimately, we can substitute equations \eqref{eq:A+A* gamma=z2square} and \eqref{eq:B+B* gamma=z2square} into equation \eqref{eq:commutator to correlator} to derive the final expression \eqref{eq:gamma=z2square}.

\subsection{Causality of non-local modular flow}\label{app:modular_flow_causality}
In this appendix, we illustrate for the case of massive scalar with $\gamma(x——2)=(x_2)^2$, that the (infinitesimally) modular flowed commutator $[[\hat{H}_\gamma,\phi(x)],\phi(y)]$ vanishes for $x\in \mathcal{R}_\gamma$ and $y\in \mathcal{L}_\gamma$. We proceed by explicitly computing the commutator in momentum space.
Since the local terms in equation \eqref{eq:gamma=z2square} remain supported in the region $\mathcal{R}_{\gamma}$,
what remains to be checked is the vanishing of the following expression,
\begin{eqnarray}\label{eq:modular flowed correlator} [[\hat{H}_{\gamma},\phi(x)],\phi(y)]
    &=& -i\pi m \left(x^-\right)^2\partial_{x^-} F(x,y)\nonumber,\\
    F(x,y)&=&\int\dd z_2\ e^{-m|x_2-z_2|}\left[\phi(x^+,x^-,z_2),\phi(y)\right].
\end{eqnarray}
Let us proceed to compute $F(x,y)$. We first express the kernel function in momentum space,
\begin{equation}
   e^{-m|x_2-z_2|}=\frac{m}{\pi}\int\dd p_2\ \frac{e^{ip_2(x_2-z_2)}}{p_2^2+m^2}.   
\end{equation}
Then we can substitute the mode expansion of the field operator and the canonical commutation relations into the definition of $F(x,y)$ in \eqref{eq:modular flowed correlator}. Integrating firstly over $z_2$ produces a factor of $\delta(p_2-q_2)$; further integrating over $p_2$  gives the following integral expression:
\begin{equation}
F(x,y)=\frac{m}{4\pi^2}\int \dd^2 q\ \frac{e^{iq\cdot (x-y)}-e^{-iq\cdot (x-y)}}{\left(q_2^2+m^2\right)E_{\vec{q}}}.
\end{equation}
where $E_{\vec{q}}=\sqrt{q_1^2+q_2^2+m^2}$ is the on-shell energy. 

This expression breaks the full Lorentz invariance but leaves the subgroup transforming only the $x^0$ and $x_1$ directions intact. We can therefore simplify it further depending on signature of $\vec{v}=\left(x^0-y^0, x_1-y_1,0\right)$ -- it is possible for $\vec{v}$ to exhibit all three signatures even though the full $x-y$ remains space-like. For space-like $\vec{v}$, one can conclude that the support of $[\hat{H}_\gamma,\phi(x)]$ lies entirely outside the light cone of $y$, as a result the $F(x,y)$ vanishes automatically. We therefore focus on the light-like and time-like $\vec{v}$. For time-like $\vec{v}$, we can perform a boost along the $x^0$ and $x_1$ direction to simplify the expression of $F(x,y)$ into:
\begin{equation}
F(x,y)=\frac{im\text{sgn}(x^0-y^0)}{2\pi^2}\int \dd^2 q\ \frac{\sin{\left(E_{\vec{q}}\Delta^{0,1}(x,y)\right)}e^{iq_2(x_2-y_2)}}{\left(q_2^2+m^2\right)E_{\vec{q}}}.
\end{equation}
where $\Delta^{0,1}(x,y)=\sqrt{(x^0-y^0)^2-(x_1-y_1)^2}>0$. One can then perform the following integral over $q_1$: 
\be
\int^\infty_{-\infty} dq_1 \frac{\sin{\left[\Delta^{0,1}(x,y)\sqrt{q_1^2+q_2^2+m^2}\right]}}{\sqrt{q_1^2+q_2^2+m^2}} = \pi J_0\left(\Delta^{0,1}(x,y)\sqrt{q_2^2+m^2}\right),
\ee
where $J_0$ is the Bessel function of the first type. We are therefore led to the following integral: 
\begin{equation}
F(x,y)=\frac{im\;\text{sgn}(x^0-y^0)}{2\pi} \int^\infty_{-\infty} dq_2 \frac{e^{iq_2(x_2-y_2)}J_0\left(\Delta^{0,1}(x,y)\sqrt{q_2^2+m^2}\right)}{q_2^2+m^2} .
\end{equation}
Apparently the integrand contains two branch points at the $q_2 = \pm im$ through the square-root in the $J_0$ function. Since $J_0(x)$ is in fact an entire function of $x^2$ -- its series expansion only contains the even powers of $x$ and has infinite radius of convergence, the branch-points are only fake. As a result, the integrand only has the two poles at $q_2 = \pm i m$ from the denominator.

We are interested in $F(x,y)$ for $x\in \mathcal{R}_\gamma$ and $y\in \mathcal{L}_\gamma$, so $x$ and $y$ are space-like separated. As a result $|x_2-y_2|>\Delta^{0,1}(x,y)$, we can complete the integration contour accordingly and pick up the residue from the corresponding pole:
\be\label{eq:time_like_F}
F(x,y)=\frac{i}{2}\text{sgn}(x^0-y^0)e^{-m|x_2-y_2|}.
\ee

The cases of light-like $\vec{v}$ proceeds similarly with simpler computations, and yields the same result as (\ref{eq:time_like_F}). We will not present the details. Combining everything, we conclude the computation with the final expression: 
\begin{eqnarray}
&&[[\hat{H}_{\gamma},\phi (x)],\phi (y)]=\frac{\pi m}{2}(x^-)^2e^{-m|x_2-y_2|}\nonumber\\
   &\times& \partial_{x^-} \left[\theta(x^+-y^+)\theta(x^--y^-)-\theta(y^+-x^+)\theta(y^--x^-)\right]   .
\end{eqnarray}

For the points of interest  $x\in \mathcal{R}_\gamma$ and $y\in \mathcal{L}_\gamma$ we have that $x^-<0,\;y^->0 \Rightarrow (y^--x^-)>0$, so the derivative operator $\partial_{x^-}$ only acts on the step function factor $\theta(y^+-x^+)$ and the commutator vanishes identically. We thus conclude that $\phi(y)$ commutes with $[\hat{H}_\gamma,\phi(x)]$ for all $y\in \mathcal{L}_\gamma$ and $x\in \mathcal{R}_\gamma$, even for those of $y$ that have causal contact with the support of the non-local part of $[\hat{H}_\gamma,\phi(x)]$. 

\bibliographystyle{JHEP} 
\bibliography{ref.bib}

\end{document}